\documentclass[amsmath, amssymb,10pt, aps, prb, twocolumn, notitlepage, longbibliography, superscriptaddress]{revtex4-2}

\usepackage{graphicx}
\usepackage{dcolumn}
\usepackage{bm}
\usepackage{hyperref}
\usepackage{xcolor}
\usepackage{braket}
\usepackage{tabularx}
\usepackage{booktabs}
\usepackage{multirow}
\usepackage{relsize}
\usepackage{bibunits}
\defaultbibliography{references}
\defaultbibliographystyle{apsrev4-2}

\newcommand{\bk}{{\bf k}}

\def\brho{{\boldsymbol \rho}}

\newcommand{\cO}{{\cal O}}

\newcommand{\vac}{\text{vac}}

\begin{document}
\begin{bibunit}[apsrev4-2]

\title{
Specific-heat anomaly in frustrated magnets with vacancy defects}    

\date{\today}

\author{Muhammad Sedik}
\affiliation{Physics Department, University of California, Santa Cruz, California 95064, USA}

\author{Siyu Zhu}
\affiliation{Physics Department, University of California, Santa Cruz, California 95064, USA}

\author{Sergey Syzranov}
\affiliation{Physics Department, University of California, Santa Cruz, California 95064, USA}

\begin{abstract}
    Motivated by frustrated magnets and spin-liquid-candidate materials, we study the thermodynamics of a 2D
    geometrically frustrated magnet with vacancy defects.
    The presence of vacancies imposes significant constraints on the bulk spins,
    which freeze some of the degrees of freedom in the system at low temperatures.
    With increasing temperature, these constraints get relaxed, resulting in an increase in the system's entropy.
    This leads to the emergence of a peak in the 
    heat capacity $C(T)$ of the magnet at a temperature $T_\text{imp}$ determined by 
    the concentration of the vacancy defects.
    The entropy associated with this peak comes from the 
    lowest-energy degrees of freedom in the material.
    To illustrate the emergence of such an anomaly,
    we compute analytically the heat capacity of
    the antiferromagnetic (AFM) Ising model on the triangular lattice with vacancy defects. The presence of the vacancy leads to a peak in $C(T)$ at the temperature 
    $T_\text{imp}=-4J/\ln n_\text{imp}$, where $J$ is the AFM coupling between
    the spins and $n_\text{imp}$ is the fraction of the missing sites.
\end{abstract}

\maketitle

Quenched disorder, such as randomly located impurities and vacancy
defects, has a profound effect on the properties of geometrically frustrated magnets (GFMs).
It can result in the emergence of the spin-glass state~\cite{BinderYoung:review}, preclude the formation of the widely sought quantum spin liquids 
(QSLs)~\cite{SavaryBalents:review}, lead to the anomalous 
``quasispin'' contribution to the magnetic susceptibility~\cite{Schiffer:TwoPopulationModel,LaForge:quasispin}
and reveal fundamental energy scales of the disorder-free material~\cite{Syzranov:HiddenEnergy,RamirezSyzranov:review}.

Despite the possible absence of long-range magnetic order in clean frustrated magnets
at low temperatures, these materials
may display slower-than-exponential decay of spin-spin correlations, making their bulk properties rather sensitive to
quenched defects and boundary conditions.

This sensitivity is exemplified by the dimer model on the hexagonal lattice, 
equivalent to the Ising model on the triangular lattice, for which the 
leading contribution to entropy at $T=0$ has been demonstrated~\cite{Elser:BoundaryDependence} to depend on the shape of the boundary, no matter how large the system is, 
in contrast with the common belief 
that the boundary conditions are not important in the thermodynamic limit.
Similar sensitivity to the boundary conditions is also known for 
the six-vertex ice model~\cite{Ferreyra:BD_one,Tavares:BD_two,Tavares:BD_three},
which may be used to simulate the behavior of certain geometrically
frustrated magnets.

This sensitivity to the boundary conditions will also result in substantial effects of quenched disorder on frustrated magnets. Such effects may be particularly strong
for vacancy defects, the most common form of quenched disorder in
GFMs, as they impose strong constraints on the bulk
degrees of freedom of magnetic materials.

In this paper, we investigate the effects of 
vacancy defects on the thermodynamics of 2D frustrated systems.
Quasi-2D materials, i.e. layered materials with weak interlayer coupling, 
are of particular interest for experimental and theoretical QSL searches.
At short length scales, the behavior of spins in such materials is effectively 2D. As there is no spin-glass transition in 2D~\cite{McMillan:firstNoTwoDGlass,HartmannYoung:TwoDGlass,Carter:TwoDGlass,Amoruso:TwoDGlass,FernandezParisi:Ising2Dtransition,Rieger:TwoDGlass} ,
quasi-2D materials tend to have very low spin-glass-freezing temperatures, set by the interlayer coupling, 
which allows them to preserve regimes favorable for potential QSL behavior
or other quantum states in a broader temperature range.

\begin{figure}[t]
    \centering
    \includegraphics[width=0.95\linewidth]{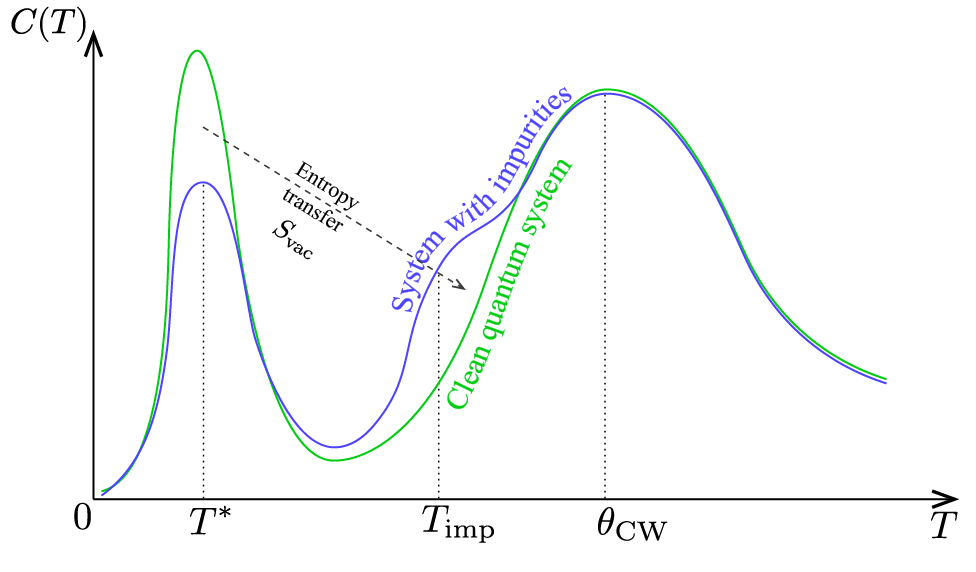}
    \caption{\label{fig:HeatCapacity} A schematic of the heat capacity of a 2D frustrated magnet in a {quantum impurity-free system (green line) and 
    in a quantum system with vacancy impurities (blue line).}
    Vacancy defects lead to an anomaly that manifests itself as a peak at a temperature
    $T_\text{imp}$ of the order of expression~\eqref{ImpurityTemperature}, which is in general
    distinct from the other temperature scales in
    the system.
    The entropy $\int_\text{peak} \left[C_\text{vac}(T)/T\right]dT
    \sim \left( N_s N_\text{imp}\right)^\frac{1}{2}$   associated with the impurity-induced peak comes predominantly from
    the low-energy states: the ground states in the case of the Ising model or the states that lead to low-temperature 
    peaks~\cite{RamirezSyzranov:review,PoppRamirezSyzranov} in Heisenberg
    models. In a generic frustrated system, the illustrated 
    heat-capacity peaks may partially merge. 
    }
\end{figure}


{\it Summary of results.} 
We show that the presence of vacancy defects in GFMs results
in an anomaly that manifests itself in the formation of a peak in the behavior of the material's heat capacity, as {illustrated} in Fig.~\ref{fig:HeatCapacity}. The temperature $T_\text{imp}$ is 
in general distinct from the other characteristic temperatures of 
the material, such as the Curie-Weiss constant $\theta_\text{CW}$.

To demonstrate the emergence of such an anomaly, we first compute the heat capacity in the Ising model focusing on the special case of the triangular
lattice.
A vacancy-free, triangular-lattice Ising model 
has degenerate ground states
with an extensive entropy with a well-known value~\cite{Wannier:Ising,Wannier:erratum,Houtappel}
of $S_0=0.323066\ldots$ per spin. The heat capacity of such a vacancy-free system per spin
\begin{align}
     C_0(T)
     =A\frac{J^2}{T^2}  e^{-\frac{4J}{T}}  
     + O \left(e^{-\frac{6J}{T}}  \right) 
     \label{HeatCapacityClean},
\end{align}
where $A=67.0622\ldots$,
is exponentially suppressed at temperatures $T\ll J$ 
significantly exceeded by the 
AFM coupling $J$ between neighbouring spins. The $4J$ activation gap is given by the 
energy of the first excited state on the triangular lattice relative to the ground-state energy~\cite{SM}. 
In this work, we recover these results, as described in Supplemental Materials~\cite{SM}, using the Kac-Ward method~\cite{kac1952combinatorial,Landafshitz5}, outlined below (with the full details presented in Supplemental Materials~\cite{SM}).

We then show that the presence of vacancy defects
in the triangular-lattice Ising model results in the
appearance of the heat-capacity peak at a temperature of the order of
\begin{align}
    T_\text{imp}=-4J/\ln n_\text{imp},
    \label{ImpurityTemperature}
\end{align}
where $n_\text{imp}=N_\text{imp}/N_s\ll 1$ is the fraction of 
missing sites in the lattice.
The contribution of the $N_\text{imp}$
vacancy defects vanishes at low temperatures
and is given by
\begin{align}
    C_\text{vac}\left(T\gtrsim T_\text{imp}\right)= \frac{2J^2N_\text{imp}}{T^2}e^\frac{2J}{T}
    \label{HeatCapacityVacancy}
\end{align}
at higher temperatures. The prefactors of $4$ and $2$
in Eqs.~\eqref{ImpurityTemperature} and \eqref{HeatCapacityVacancy}
are specific to the triangular lattice and will differ by factors of order
unity for Ising models on other 2D frustrating lattices.

In general, the vacancy-induced peak will appear below the Curie-Weiss temperature
$\theta_\text{CW}\sim z J$ (see Fig.~\ref{fig:HeatCapacity}), a characteristic energy scale of spin-flip-type excitations, where $z$ is the coordination number.
The entropy $S_\text{vac}=\int_\text{peak} \frac{C_\text{vac}(T)}{T}dT \simeq \left(N_s N_\text{imp}\right)^\frac{1}{2}$
associated
with the peak comes predominantly from the ground states of the Ising model (hereinafter, when discussing the scaling of contributions to the heat capacity as functions of $N_\text{imp}$, we neglect slow logarithmic factors $\propto \ln n_\text{imp}$). 
While the presence of the vacancies decreases the total entropy 
$\int_0^\infty \frac{C(T)}{T}dT$ 
by $N_\text{imp}\ln 2$, this decrease is significantly smaller than the entropy 
$S_\vac$
associated with the vacancy-induced peak. The main effect of vacancy defects is, therefore, redistributing the entropy between the degrees of freedom with low energies and with energies of order $T_\text{imp}$, creating a peak at the latter temperature.
The described effects are not specific to Ising models and 
are observable in rather generic quantum GF magnets.

The vacancy-induced peak
at temperature $T\sim T_\text{imp}$
persists for higher spins $s$ in the presence of the transverse spin-spin
coupling, i.e. in Heisenberg models on GF lattices. 
Such GF materials have 
recently been demonstrated~\cite{PoppRamirezSyzranov}
to generically exhibit
a two-peak structure in the temperature 
dependence $C(T)$ of the heat capacity (see Fig.~\ref{fig:HeatCapacity})
in the absence of impurities. Vacancy impurities lead to the emergence
of an impurity peak with the entropy 
at a temperature $T\sim T_\text{imp}$ of the order of expression~\eqref{ImpurityTemperature},
which in general is distinct from the 
characteristic temperatures $T^*$
and $\theta_\textit{CW}$ of the other peaks.

\begin{figure}[t!]
	\centering
	\includegraphics[width=0.9\linewidth]{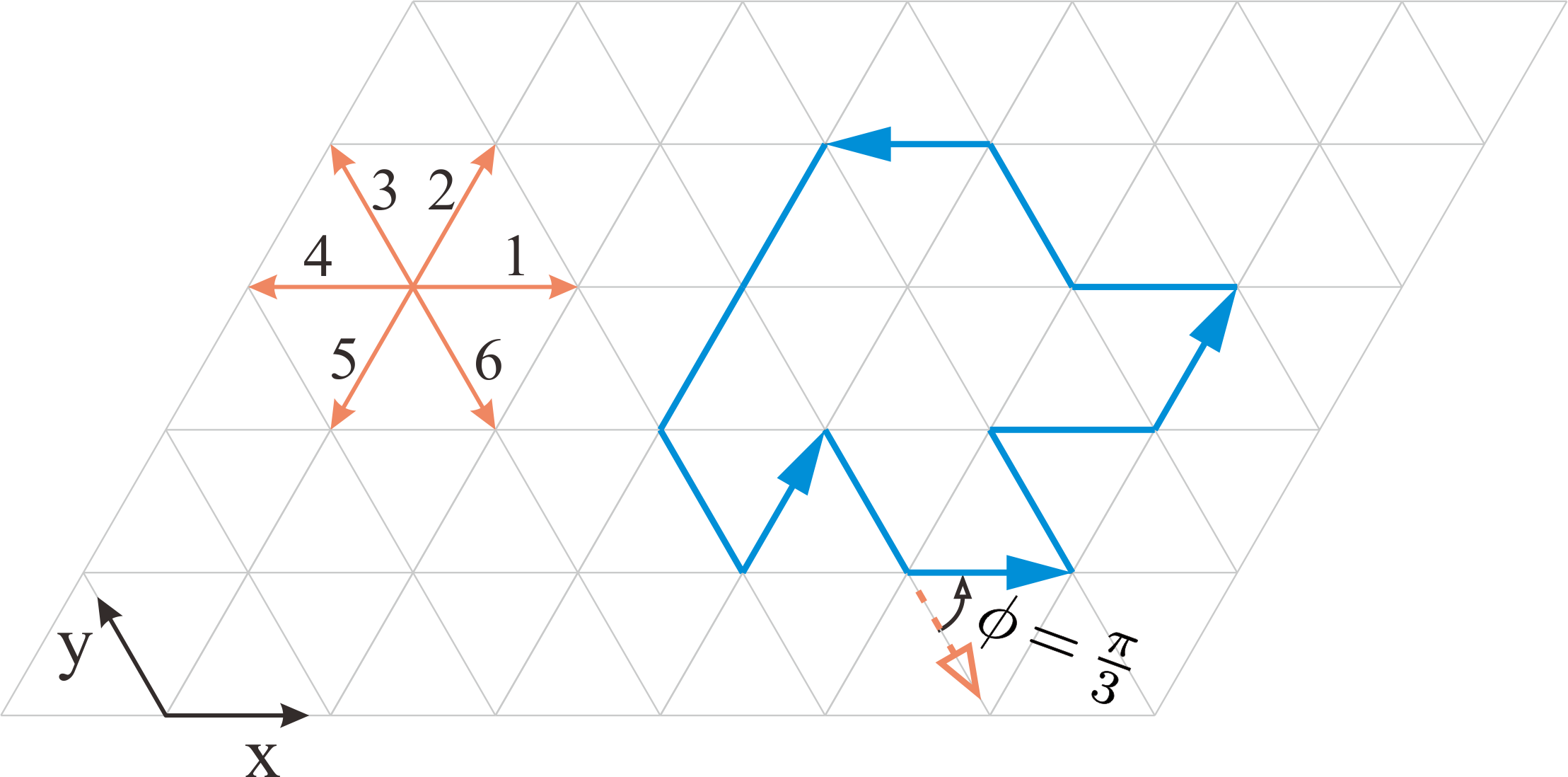}
	\caption{\label{fig:loop} 
    A closed loop of bonds of length $r=12$ on the triangular lattice. 
    At each site, the direction of the loop rotates by angle $\phi$,
    which is taken into account when evaluating the contribution of the loop
    to the partition function~\eqref{PartitionLoop}.
    The orange arrows show six possible directions for each bond of the loop.
    }
\end{figure}

{\it Ising partition function as a sum over 2D loops.} 
In what immediately follows,
we describe the heat capacity derivation
in the clean Ising model and in that with vacancy defects.
The partition function of a 2D system of $N_s$ Ising spins connected by $N_b$
links with the antiferromagnetic coupling $J$ can be expressed
in the form~\cite{kac1952combinatorial,Landafshitz5}
\begin{align}
    Z=2^{N_s}\left(\cosh\frac{J}{T}\right)^{N_b}
    \exp\left[-\sum_{r=1}^{\infty}\left(-\tanh\frac{J}{T}\right)^r f_r \right],
    \label{PartitionLoop}
\end{align}
where 
$f_r$ is the sum of 
the factors $(-1)^\zeta$ for all closed loops of length $r$,
as shown in Fig.~\ref{fig:loop},
with $\zeta$ being the number of self-intersections for a particular loop; hereinafter 
$k_B=1$.

Evaluating the sum $f_r$ over all possible loops is reduced to the problem of random walk in the space of the lattice sites $\brho_i$ and the direction
$\nu_i$ of the loop at each respective site, where for the triangular lattice, $\nu_i$ 
labels one of six possible directions (see Fig.~\ref{fig:loop}).
The parity $(-1)^\zeta$ of each loop can be taken into account by assigning 
a factor of $\exp(i\phi/2)$ at each site of the loop, where $\phi$
is the angle of rotation of the tangent to the loop at that site, as shown 
in Fig.~\ref{fig:loop}.

The contribution of each loop can be expressed as a product of the phase factors: 
\begin{align}
    &f_r=
    \nonumber\\
    &\frac{1}{2r}
    \sum_{\substack{\brho_1 ,\brho_2\ldots,\brho_r 
    \\ 
    \nu_1,\nu_2,\ldots,\nu_r}}
    \Lambda\left(\brho_1\nu_1|\brho_2\nu_2\right)
    \Lambda\left(\brho_2\nu_2|\brho_3\nu_3\right)
    \ldots
    \Lambda\left(\brho_r\nu_r|\brho_1\nu_1\right),
    \label{fr}
\end{align}
where $\Lambda\left(\brho_i\nu_i|\brho_{i+1}\nu_{i+1}\right)=
\exp\left(i\phi_i/2\right)$ is the phase that the loop acquires
between nearest-neighbour sites $\brho_i$ and $\brho_{i+1}$,
provided the directions $\nu_i$ and $\nu_{i+1}$ are not opposite to each other.
If the random walk makes a U-turn, we require that 
$\Lambda\left(\brho_i\nu_i|\brho_{i+1}\nu_{i+1}\right)=0$ to disallow 
walking backwards along the same 
bond, as the contribution of each bond in Eq.~\eqref{PartitionLoop}
should be counted only once~\cite{Footnote1}.
The $\frac{1}{2r}$ prefactor in Eq.~\eqref{fr}
prevents overcounting loops due to two possible directions 
and the possibility of the starting point $\brho_1$ lying anywhere on 
the same loop of $r$ sites.

{\it Thermodynamics of the clean system.}
In what immediately follows, we 
use the loop representation of the partition function~\eqref{PartitionLoop}
to reproduce the well-known result for the zero-temperature limit 
of the Ising model on the triangular lattice, which was obtained 
in Refs.~\cite{Wannier:Ising,Wannier:erratum} and \cite{Houtappel}
using other methods.

In the absence of impurities, the system is translationally invariant, and the
contribution~\eqref{fr}
of loops of length $r$ can be conveniently evaluated in the momentum representation. Writing down explicitly the element $\Lambda\left(\brho\nu|\brho^\prime\nu^\prime\right)$ for all
possible mutual positions $\brho-\brho^\prime$ and all possible directions 
$\nu$ and $\nu^\prime$ and performing the Fourier-transform with respect to the
coordinate difference $\brho-\brho^\prime$ {(see the Supplemental Material~\cite{SM} for details)} gives a $6\times6$ matrix in the 
direction space:
\begin{widetext}
\begin{align}
	{\Lambda}_\bk
	&=
	\begin{pmatrix}
		\epsilon^{-k_x} & \alpha^{-1}\epsilon^{-k_x}\epsilon^{-k_y} & \alpha^{-2}\epsilon^{-k_y} & 0 & \alpha^{2}\epsilon^{k_x}\epsilon^{k_y} & \alpha\,\epsilon^{k_y}\\
		\alpha\,\epsilon^{-k_x} & \epsilon^{-k_x}\epsilon^{-k_y} & \alpha^{-1}\epsilon^{-k_y} & \alpha^{-2}\epsilon^{k_x} & 0 & \alpha^{2}\epsilon^{k_y}\\
		\alpha^{2}\epsilon^{-k_x} & \alpha\,\epsilon^{-k_x}\epsilon^{-k_y} & \epsilon^{-k_y} & \alpha^{-1}\epsilon^{k_x} & \alpha^{-2}\epsilon^{k_x}\epsilon^{k_y} & 0\\
		0 & \alpha^{2}\epsilon^{-k_x}\epsilon^{-k_y} & \alpha\,\epsilon^{-k_y} & \epsilon^{k_x} & \alpha^{-1}\epsilon^{k_x}\epsilon^{k_y} & \alpha^{-2}\epsilon^{k_y}\\
		\alpha^{-2}\epsilon^{-k_x} & 0 & \alpha^{2}\epsilon^{-k_y} & \alpha\,\epsilon^{k_x} & \epsilon^{k_x}\epsilon^{k_y} & \alpha^{-1}\epsilon^{k_y}\\
		\alpha^{-1}\epsilon^{-k_x} & \alpha^{-2}\epsilon^{-k_x}\epsilon^{-k_y} & 0 & \alpha^{2}\epsilon^{k_x} & \alpha\,\epsilon^{k_x}\epsilon^{k_y} & \epsilon^{k_y}
	\end{pmatrix},
        \label{LambdaFT}
\end{align}
\end{widetext}
where $\bk=(k_x,k_y)$ is the 2D momentum conjugate to the position difference $\brho-\brho^\prime$ for the coordinate system shown in Fig.~\ref{fig:loop}; $\alpha = \exp\left(\,i\pi/6\right)$; $\epsilon = \exp\left({2\pi i}/{L}\right)$, and $L=N_s^\frac{1}{2}$ is the linear size
of the system (measured in units of the lattice spacing) in both $x$ and $y$ directions (cf. Fig.~\ref{fig:loop}).

Utilising Eqs.~\eqref{PartitionLoop} and \eqref{fr}, the partition function of the defect-free system can be related to the matrix $\Lambda_\bk$ as
\begin{align}
    &Z_0=
    \nonumber\\
    &2^{N_s}\left(\cosh\frac{J}{T}\right)^{N_b}
    \exp \left[ - \sum_\bk \sum_{r=1}^{\infty} 
    \frac{1}{2r} \left(-\tanh\frac{J}{T}\right)^r 
    \operatorname{Tr}\Lambda_\bk^r \right].
    \label{ZviaLambda}
\end{align}
Substituting the matrix~\eqref{LambdaFT} into Eq.~\eqref{ZviaLambda}, we obtain the entropy of the system per spin in the form
(see Supplemental Material~\cite{SM} for the details of the calculations)
\begin{align}
    S_0(T) = S_0 \left(0\right) +\frac{A}{16}\left(1+\frac{4J}{T}\right) e^{-\frac{4J}{T}}  + O \left(e^{-\frac{6J}{T}}  \right) ,
    \label{EntropyVacancyResult}
\end{align}
and the heat capacity~\eqref{HeatCapacityClean} of the defect-free system,
where 
\begin{align}
    &S_0 \left(0\right)
    = 
    \nonumber\\
    &\frac{1}{8\pi^2} \int_0^{2\pi} \int_0^{2\pi} \ln \left( 1 - 4 \cos \omega \cos\omega' + 4\cos^2\omega' \right) \, d\omega \, d\omega'
    \nonumber\\
    & 
    =  0.323066\ldots
    \label{ZeroPointEntropyClean}
\end{align}
is the entropy in the limit of vanishing temperature~\cite{Wannier:Ising},
and the coefficient $A$ in Eq.~\eqref{EntropyVacancyResult}
is defined after Eq.~\eqref{HeatCapacityClean}.
The value~\eqref{ZeroPointEntropyClean} of the entropy of the clean material 
matches Wannier's result~\cite{Wannier:Ising,Wannier:erratum,Houtappel}.

{\it Contribution of a single vacancy.} 
Next, we proceed to the case of an Ising
system with a single vacancy
defect at a certain location $\brho_0$.
The partition function $Z$ of such a system can be computed using the loop representation~\eqref{PartitionLoop}.
In the presence of a vacancy at site $\brho_0$,
the loops that contribute to $Z$ do not contain site $\brho_0$.

The partition function of the system with a vacancy is given by
\begin{align}
    Z= 2^{-1} \left(\cosh\frac{J}{T}\right)^{-6}
    \exp\left[\sum_{r=1}^{\infty}\left(-\tanh\frac{J}{T}\right)^r \tilde f_r \right] \,Z_0,
    \label{PartitionFunctionWithVacancy}
\end{align}
where $Z_0$ is the partition function of the vacancy-free system;
the prefactors $2^{-1}$ and $\left(\cosh\frac{J}{T}\right)^{-6}$
reflect one missing site and six missing links in the system with a vacancy;
$\tilde{f}_r$ is the sum over all the loops that include site $\brho_0$.
The contribution of such loops is subtracted from the contribution of all possible loops
in the partition function $Z_0$, thus leaving the contributions 
of loops that do not contain site $\brho_0$.

Because the thermodynamic functions in a sufficiently large system 
do not depend on the location of the vacancy, 
a thermodynamic observable $\cO$ can be averaged with respect to the location $\brho_0$
of the vacancy site as $\cO=\left<\cO\right>\equiv\frac{1}{N_s}\sum_{\brho_0}\cO$.
Such averaging is equivalent to the replacement
$\tilde{f}_r\rightarrow \frac{r}{N_s} f_r$ in the partition function~\eqref{PartitionFunctionWithVacancy},
where $f_r$ is the sum over loops in the clean system
and the prefactor $r$ reflects that before averaging; each loop corresponds to
$r$ possible locations of site $\brho_0$.

The series in Eq.~\eqref{PartitionFunctionWithVacancy} averaged with respect to $\brho_0$ can be computed by differentiating
a similar series in Eq.~\eqref{PartitionLoop} with respect to $\tanh\frac{J}{T}$.
We find~\cite{SM} the entropy of the system with a vacancy to be given by $S=S_0 + S_1$,
where $S_0$
is the entropy of the vacancy-free system, and
the vacancy contribution $S_1$ in the limit of low temperatures $T\ll J$ is
given by
\begin{align}
     S_1\approx -\frac{J}{T} e^{\frac{2J}{T}}.
     \label{SingleVacancyEntropy}
\end{align}

\begin{figure}[t!]
	\includegraphics[width=0.95\linewidth]{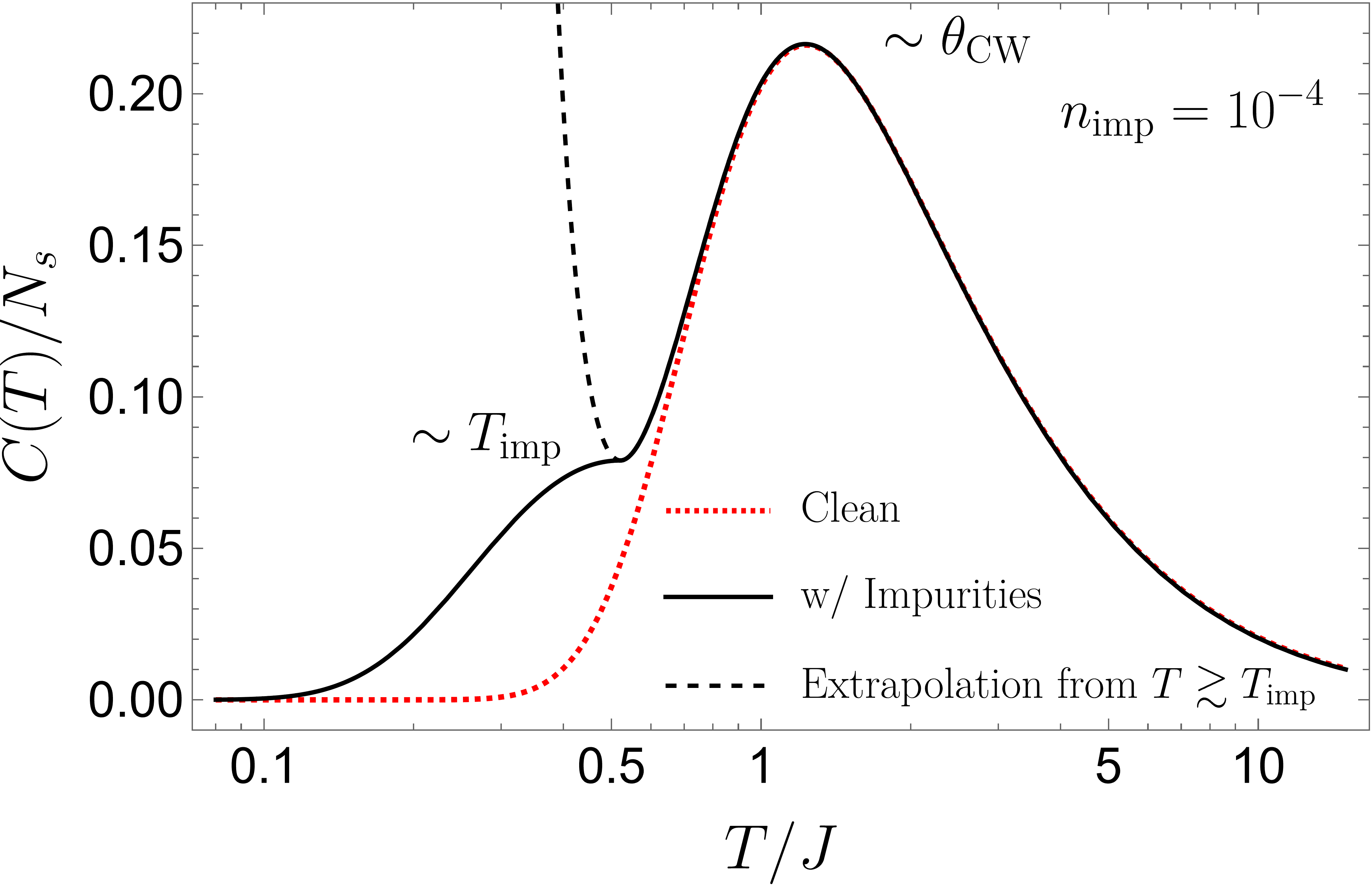}
	\caption{
    \label{fig:Ising_heat_capacity} 
    Heat capacity per site for an impurity-free Ising system on the triangular lattice (dotted red) and a system with impurities. The black dashed line shows the extrapolation of the analytical result valid for $T\gtrsim T_{\text{imp}}$ to lower temperatures. The black solid line is the total heat capacity obtained by interpolating between the exact result for $C(T)$ at $T\gtrsim T_\text{imp}$ and $C(T)=0$ at $T=0$.     
    }
\end{figure}

{\it Heat capacity of multiple vacancies.}
In a system with multiple vacancies, the interplay between different vacancies can be neglected at 
sufficiently high temperatures $T\gtrsim T_\text{imp}\sim -J/\ln n_\text{imp}$, at which
the characteristic length~\cite{Jacobsen:correlationLength}
$\xi(T)=\exp\left(2J/T\right)$ of spin correlations (measured in the units of lattice spacing)
is significantly smaller than the 
typical distance $n_\text{imp}^{-\frac{1}{2}}$ between vacancies.
At such temperatures, the contributions of different vacancies to the 
entropy and heat capacity are additive, and the contribution of vacancies to 
the heat capacity is given by Eq.~\eqref{HeatCapacityVacancy}, as follows
from Eq.~\eqref{SingleVacancyEntropy}.
The resulting heat capacity for the Ising model on the triangular lattice is shown with a dashed black line in Fig.~\ref{fig:Ising_heat_capacity}.

The growth of the heat capacity~\eqref{HeatCapacityVacancy} with decreasing the temperature $T$ persists down to the temperature $T_\text{imp}$,
at which the correlation length $\xi(T)$ reaches the typical inter-vacancy
distance $n_\text{imp}^{-\frac{1}{2}}$ {(see Supplemental Material~\cite{SM} for the derivation details)}.
At low temperatures, the heat capacity vanishes, $C(T\rightarrow0)\rightarrow0$,
as required by the convergence of the entropy $S(T)=S_0(0)+\int_0^T\frac{C(T)}{T}dT$.
The positive contribution to the heat capacity at $T\lesssim T_\text{imp}$ that grows with lowering temperature and the vanishing $C(T)=0$ at $T=0$ requires the existence of an impurity-induced $C(T)$ anomaly (peak), as shown by the black line in Fig.~\ref{fig:Ising_heat_capacity}.
While we do not present an exact analytical calculation of the 
heat capacity $C(T)$ at temperatures $0<T\lesssim T_\text{imp}$, the existence of the peak follows from the discussed $C(T)$ asymptotics outside of this interval.

\begin{figure}[t!]
	\includegraphics[width=0.9\linewidth]{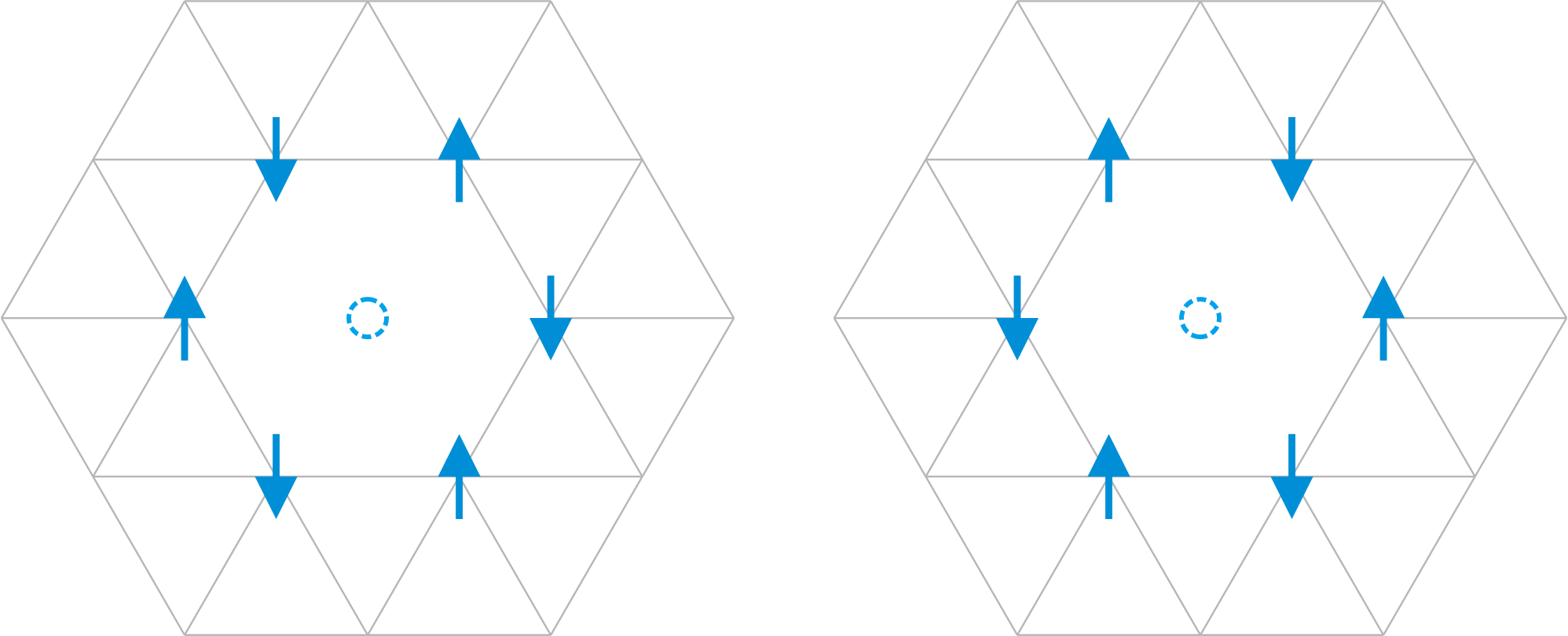}
	\caption{
    \label{fig:vacancy} 
    Spin configuration around a vacancy in a triangular lattice showing the 
    alternating pattern at the sites next to the vacancy. 
    }
\end{figure}

{\it Qualitative interpretation of the vacancy-induced peak.} 
At $T=0$, vacancies significantly constrain the bulk degrees of freedom
in a frustrated magnet. For example, for the Ising model on the triangular lattice, only spin
patterns with staggered magnetisation around the vacancy are allowed
at low temperatures, as shown in Fig.~\ref{fig:vacancy},
which restricts the set of possible ground states relative to the case 
of the clean system.
The constraint imposed by a vacancy on the bulk degrees of freedom 
is qualitatively similar to that imposed by a boundary, which is known to lower the 
bulk entropy of an arbitrarily large frustrated magnet at $T=0$~\cite{Elser:BoundaryDependence,MillaneBlakeley:boundaryDependence,Destainville:boundaryDependence}.

At low temperatures, the constraints imposed by the vacancy
defects persist so long as the correlation length $\xi(T)$
of the clean frustrated medium exceeds the inter-vacancy distance $n_\text{imp}^{-1/2}$ [for Ising models, $\xi(T)\sim \exp\left(aJ/T\right)$, with $a=2$ for the triangular lattice~\cite{Jacobsen:correlationLength}].
At temperature $T\sim T_\text{imp}$, these lengths are of the same order of magnitude. With further increasing the temperature, 
the decrease of the correlation $\xi(T)$ length 
leads to a rapid shrinking of the constrained regions around
individual vacancies.
The resulting growth of entropy at $T\sim T_\text{imp}$
gives a peak in the heat capacity.

{\it Other characteristic temperature scales and the heat-capacity structure.}
The described mechanism of the 
emergence of a heat-capacity peak at $T\sim T_\text{imp}$ is not specific to Ising systems and persists also in GF magnets described by the Heisenberg model. 
In such models with the ratio of the transverse-to-longitudinal coupling $J_{xx}/J_{zz}\lesssim 1$, the transverse coupling hybridises the Ising ground states, which results in a distinct heat-capacity peak with the characteristic temperature $T^*\lesssim \theta_\text{CW}$ (see Fig.~\ref{fig:HeatCapacity}), which may be well separated from the Curie-Weiss peak or partially merge with it (see Refs.~\cite{RamirezSyzranov:review} and \cite{PoppRamirezSyzranov} for a review).
In a {\it clean} GF magnet, the heat capacity, thus,
displays a two-peak structure~\cite{Greywall:DoublePeakHe,Schiffer:GGG,Ishida:HeRingExchange,NakatshujiMaeno:NaGaSneutron,Li:S2TAF,Bordelon:NaYbO2,Ranjith:NaYbSe2}.

The presence of vacancy defects modifies this structure, giving rise to a third peak at $T\sim T_\text{imp}$, as shown in Fig.~\ref{fig:HeatCapacity}, which
is readily distinguishable from the peaks 
at $T\sim T^*$ and $T\sim\theta_\text{CW}$ for realistic impurity concentrations.
The impurity peak may be masked in very clean samples 
or in samples where the low-temperature peak at $T\sim T^*$ is weakly distinguishable from
the Curie-Weiss peak or forms a ``shoulder'' in its
vicinity~\cite{SugiuraShimizu:KHAF,Munehisa:KHAF,ChenQu:THAF,PrelovsekKokalj:THAF,SchnackSchulenberg:KHAF,Gonzalez:THAF}.

{\it Entropy balance and heat-capacity peaks.} 
Because the presence of vacancies increases the heat capacity $C(T)$ near the temperature $T_\text{imp}$, it also increases
the magnetic entropy $S(T)=\int_\text{peak} C(T)/T\, dT$ released by the material when cooled down in the respective temperature interval by an amount of the order 
$|S_\text{vac}(T_\text{imp})|\sim \left(N_s N_\text{imp}\right)^\frac{1}{2}$. 
In an Ising system,
this entropy comes from the ground states, as vacancies reduce 
the ground–state entropy per spin by partially
lifting the clean system's degeneracy $T\!\to\!0$.
In a quantum magnet, this entropy comes from the entropy of the lower-temperature peak at temperature  $T\sim T^*$, as shown in
Fig.~\ref{fig:HeatCapacity} (here, we neglect the small decrease
of the total entropy $\int_0^\infty \frac{C(T)}{T}dT$ by the amount
$N_\text{imp}\ln 2$ caused by the presence of the vacancies).


{\it Conclusion.}
We have examined the effects of vacancy defects on the thermodynamics 
of frustrated magnets focussing on the Ising model on the triangular lattice. 
At low temperatures, vacancy defects constrain and effectively freeze some of the degrees of freedom in the system.
These constraints get relaxed with increasing temperature, which leads to 
a peak in the heat capacity of the system at low temperatures determined by the vacancy concentration.

We illustrate the emergence of such a vacancy-induced low-temperature 
anomaly by computing analytically the heat capacity in the Ising model on
the triangular lattice with vacancy defects.

{\it Acknowledgements.} We are indebted to A.P.~Ramirez for 
useful discussions and feedback on the manuscript. We also thank Phillip Popp for useful discussions and bringing Refs.~\cite{Landau:Vortices} and \cite{Moore:LogInteractions}. This work has been supported by the NSF grant DMR2218130.


\putbib[references]
\end{bibunit}



\begin{bibunit}[apsrev4-2]
\newpage
\onecolumngrid
\vspace{2cm}

\cleardoublepage

\setcounter{page}{1}

\renewcommand{\theequation}{S\arabic{equation}}
\renewcommand{\thefigure}{S\arabic{figure}}
\renewcommand{\thetable}{S\arabic{table}}
\renewcommand{\bibnumfmt}[1]{[S#1]}
\renewcommand{\citenumfont}[1]{S#1}

\setcounter{equation}{0}
\setcounter{figure}{0}
\setcounter{enumiv}{0}

\begin{center}
	\textbf{\large Supplemental Material for \\
        ``Specific-heat anomaly in frustrated magnets with vacancy defects"
	}
\end{center}

\section{Thermodynamics of the defect-free system}
\label{SuppSec:vacancy-free}

In this section, we derive the partition function of the defect-free
Ising model on the
triangular lattice. Our approach was first introduced by Kac and Ward~\cite{kac1952combinatorial}, and it follows the pedagogical description in Landau and Lifshitz's book~\cite{Landafshitz5} for the square lattice.
The energy of a generic 2D Ising model with nearest-neighbor
interactions is given by
\begin{align}
	E(\sigma) = J \sum_{\langle (x, y),(x', y') \rangle}  \sigma_{x,y} \sigma_{x',y'}, \label{}
\end{align}
where the sites are labelled by their 2D coordinates $(x,y)$ and
the sum $\sum_{\langle (x, y),(x', y') \rangle}$ runs over all nearest-neighbor pairs.
The partition function is given by
\begin{align}
	Z_0 = \sum_{\{\sigma\}} e^{-E(\sigma)/T} = \sum_{\{\sigma\}} \exp \left( -\frac{J}{T} \sum_{\langle (x, y),(x', y') \rangle} \sigma_{x,y} \sigma_{x',y'} \right). \label{}
\end{align}
By utilizing the identity
\begin{align}
	e^{-\frac{J}{T} \sigma_{x,y} \sigma_{x',y'}} = \cosh \frac{J}{T} - \sigma_{x,y} \sigma_{x',y'} \sinh \frac{J}{T} = \cosh \frac{J}{T} \left( 1 - \sigma_{x,y} \sigma_{x',y'} \tanh \frac{J}{T} \right), \label{}
\end{align}
the partition function can be rewritten as
\begin{align}
	Z_0 = \left(1 - t^2\right)^{-N_b/2} R, \label{eq:Z_in_R}
\end{align}
where $N_b$ is the number of bonds, and 
we have defined the quantity
\begin{align}
	R \equiv \sum_{\{\sigma\}} \prod_{\langle (x, y),(x', y') \rangle} \left( 1 + t\, \sigma_{x,y} \sigma_{x',y'} \right), \label{}
\end{align}
as well as
\begin{align}
    t=-\tanh \frac{J}{T}.   
    \label{tDefinition}
\end{align}

Since $R$ is a polynomial in terms of $t$ and $\sigma$, each term in the expansion corresponds to a specific set of bonds connecting adjacent lattice points. Because $\sigma_{xy}=\pm 1$, the summation over all configurations cancels all the terms where any $\sigma_{xy}$ appears an odd number of times. Therefore, only those terms with even powers of $\sigma_{xy}$ remain,
which implies that the nonvanishing contributions to the partition function come only from closed graphs on the lattice.

Carrying out the summation over all the spins $\sigma$, the function $R$
can be represented in terms loop diagrams on the given lattice as
\begin{align}
	R = 2^{N_s} \sum_r t^r g_r, \label{}
\end{align}
where the factor $2^{N_s}$ represents the total number of spin configurations, and $g_r$ is the number of sets of closed loops, with $r$
being the number of bonds in each set.
Following the approach of Ref.~\cite{Landafshitz5}, the function $R$ can be rewritten as
\begin{align}
	R = 2^{N_s} \exp \left( -\sum_{r=1}^\infty t^r f_r \right),
    \label{eq:R_0}
\end{align}
where $f_r$ is the sum of parities $(-1)^\zeta$ 
of all closed loops of length $r$, in which $\zeta$ is the number of self-intersections for a particular loop.

We note that Eq.~\eqref{eq:R_0}
applies to an arbitrary lattice.
In what follows we specialise to the triangular lattice, on which the quantity $R$ can be obtained by analysing a weighted random walk.

\begin{figure}[h]
	\centering
	\includegraphics[width=0.5\textwidth]{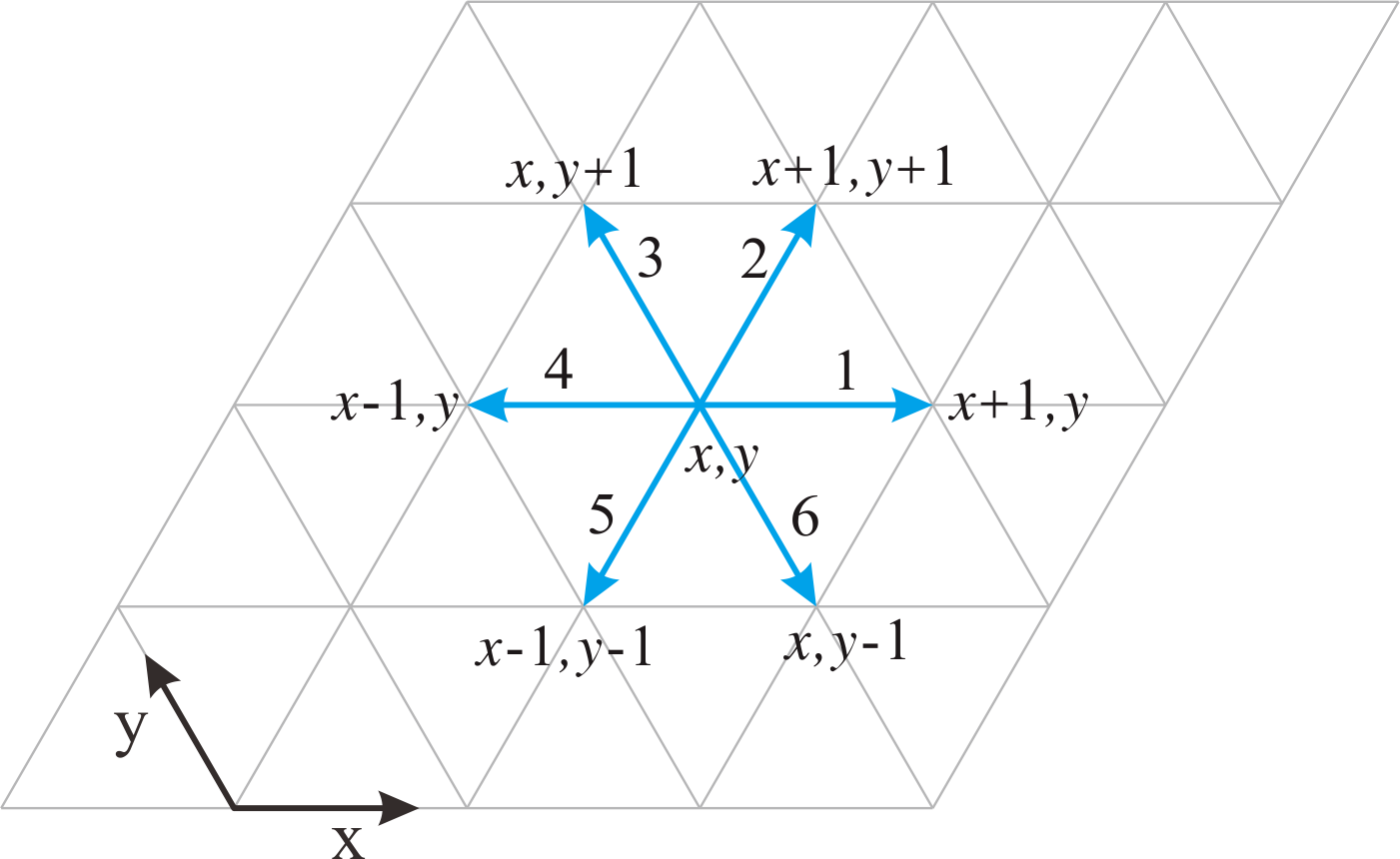} 
	\caption{Nearest neighbor position relationship in the triangular lattice.
	}
	\label{fig:position}
\end{figure}

We label the six nearest‑neighbour directions of the triangular lattice by the index $\nu\in\{1,\dots,6\}$, as illustrated in Fig.~\ref{fig:position}.
Each step of the walk carries a phase factor $e^{i\phi/2}$, where $\phi$ is the rotation angle at that vertex.  Hereinafter,
all the coordinates are given in the skew coordinate system on the triangular lattice,
as shown in Fig.~\ref{fig:position}.

We define the sum
\begin{align}
    W_r(x,y,\nu)
    \equiv
    \sum_{\text{paths of length }r}
    e^{\frac{i}{2}\sum \phi}
    \label{Wdefinition}
\end{align}
over all paths that start at the location $(x_0,y_0)$ with the direction $\nu_0$
and end at the site $(x,y)$ with direction \textbf{not} from the point to which the arrow $\nu$ directed. 
Accordingly, the quantity $W_r(x_0,y_0,\nu_0)$ gives the sum of weighted closed loops 
that contain the site $(x_0,y_0)$ and have the direction $\nu_0$ at location $(x_0,y_0)$.
Because each loop can be traversed in two directions 
and contain $r$ lattice sites,
\begin{align}
	f_r = \frac{1}{2r} \sum_{x_0, y_0, \nu_0} W_r(x_0, y_0, \nu_0). 
    \label{f}
\end{align}

The sum $W_{r+1}(x,y,\nu)$ of the weighted paths of length $r+1$ can be related to a similar 
sum $W_{r}(x',y',\nu')$ over paths of lengths $r$ as
\begin{align}
	W_{r+1}(x,y,\nu) = \sum_{x',y',\nu'} \Lambda (xy\nu \mid x'y'\nu')\, W_r(x',y',\nu'), \label{RecurrenceMatrixDefinition}
\end{align}
where the recurrence matrix $\Lambda(xy\nu \mid x'y'\nu')$ describes the propagation from the nearest neighboring site $(x',y')$ and direction $\nu'$ to the site $(x,y)$ and $\nu$.
Using the direction labels shown in 
Fig.~\ref{fig:position} and considering the change
of the phase $\phi$ [cf. Eq.~\eqref{Wdefinition}] for the links of the path along the 
respective directions, we rewrite Eq.~\eqref{RecurrenceMatrixDefinition} in the form
\begin{align}
	\begin{pmatrix}
		W_{r+1}(x,y,1)\\
		W_{r+1}(x,y,2)\\
		W_{r+1}(x,y,3)\\
		W_{r+1}(x,y,4)\\
		W_{r+1}(x,y,5)\\
		W_{r+1}(x,y,6)
	\end{pmatrix}
	&=
	\begin{pmatrix}
		1 & e^{-\frac{i\pi}{6}} & e^{-\frac{i\pi}{3}} & 0 & e^{\frac{i\pi}{3}} & e^{\frac{i\pi}{6}}\\
		e^{\frac{i\pi}{6}} & 1 & e^{-\frac{i\pi}{6}} & e^{-\frac{i\pi}{3}} & 0 & e^{\frac{i\pi}{3}}\\
		e^{\frac{i\pi}{3}} & e^{\frac{i\pi}{6}} & 1 & e^{-\frac{i\pi}{6}} & e^{-\frac{i\pi}{3}} & 0\\
		0 & e^{\frac{i\pi}{3}} & e^{\frac{i\pi}{6}} & 1 & e^{-\frac{i\pi}{6}} & e^{-\frac{i\pi}{3}}\\
		e^{-\frac{i\pi}{3}} & 0 & e^{\frac{i\pi}{3}} & e^{\frac{i\pi}{6}} & 1 & e^{-\frac{i\pi}{6}}\\
		e^{-\frac{i\pi}{6}} & e^{-\frac{i\pi}{3}} & 0 & e^{\frac{i\pi}{3}} & e^{\frac{i\pi}{6}} & 1
	\end{pmatrix}
	\begin{pmatrix}
		W_{r}(x-1,y,1)\\
		W_{r}(x-1,y-1,2)\\
		W_{r}(x,y-1,3)\\
		W_{r}(x+1,y,4)\\
		W_{r}(x+1,y+1,5)\\
		W_{r}(x,y+1,6)
	\end{pmatrix}. 
        \label{eq:W}
\end{align}



Summation over all paths of length $r$ gives
\begin{align}
	\text{tr}\, \Lambda^r = \sum_{x_0,y_0,\nu_0} W_r(x_0,y_0,\nu_0), \label{}
\end{align}
where $\Lambda (xy\nu \mid x'y'\nu')$ is the recurrence matrix given by Eq.~\eqref{RecurrenceMatrixDefinition}.
The quantity $f_r$ defined by Eq.~\eqref{f}, can be expressed through the eigenvalues $\lambda_i$
of the matrix $\Lambda (xy\nu \mid x'y'\nu')$ as
\begin{align}
	f_r = \frac{1}{2r} \text{tr}\, \Lambda^r = \frac{1}{2r} \sum_i \lambda_i^r.
    \label{frThroughLambdaI}
\end{align}

Substituting Eq.~\eqref{frThroughLambdaI} into
Eq.~\eqref{eq:R_0} and interchanging the order of summation with respect to $i$ and $r$ gives
\begin{align}
	R =  2^{N_s}  \exp \left( -\frac{1}{2} \sum_i \sum_{r=1}^{\infty} \frac{1}{r} t^r \lambda_i^r \right) 
	=  2^{N_s}  \exp \left[ \frac{1}{2} \sum_i \ln(1 - t \lambda_i) \right] 
	=  2^{N_s}  \prod_i \sqrt{1 - t \lambda_i}. 
    \label{R_1}
\end{align}

The eigenvalues $\lambda_i$ can be computed by Fourier-transforming the 
recursion relation~\eqref{eq:W}.
%
%
Introducing the Fourier transform 
\begin{align}
	\tilde{W}_{r}(k_x,k_y,\nu)
	=
	\sum_{x=0}^{L-1}\sum_{y=0}^{L-1}
	e^{-\frac{2\pi i}{L}(k_x\,x+k_y\,y)}\,W_{r}(x,y,\nu), 
        \label{}
\end{align}
of the $\Lambda (xy\nu \mid x'y'\nu')$ matrix, where $\nu=1,\dots,6$ and $L\times L$ is the 
size of the system (measured in elementary cells), Eq.~\eqref{eq:W} can be recast in the form
\begin{align}
	\begin{pmatrix}
		\tilde{W}_{r+1}(k_x,k_y,1)\\
		\tilde{W}_{r+1}(k_x,k_y,2)\\
		\tilde{W}_{r+1}(k_x,k_y,3)\\
		\tilde{W}_{r+1}(k_x,k_y,4)\\
		\tilde{W}_{r+1}(k_x,k_y,5)\\
		\tilde{W}_{r+1}(k_x,k_y,6)
	\end{pmatrix}
	&=
        \begin{pmatrix}
		\epsilon^{-k_x} & \alpha^{-1}\epsilon^{-k_x}\epsilon^{-k_y} & \alpha^{-2}\epsilon^{-k_y} & 0 & \alpha^{2}\epsilon^{k_x}\epsilon^{k_y} & \alpha\,\epsilon^{k_y}\\
		\alpha\,\epsilon^{-k_x} & \epsilon^{-k_x}\epsilon^{-k_y} & \alpha^{-1}\epsilon^{-k_y} & \alpha^{-2}\epsilon^{k_x} & 0 & \alpha^{2}\epsilon^{k_y}\\
		\alpha^{2}\epsilon^{-k_x} & \alpha\,\epsilon^{-k_x}\epsilon^{-k_y} & \epsilon^{-k_y} & \alpha^{-1}\epsilon^{k_x} & \alpha^{-2}\epsilon^{k_x}\epsilon^{k_y} & 0\\
		0 & \alpha^{2}\epsilon^{-k_x}\epsilon^{-k_y} & \alpha\,\epsilon^{-k_y} & \epsilon^{k_x} & \alpha^{-1}\epsilon^{k_x}\epsilon^{k_y} & \alpha^{-2}\epsilon^{k_y}\\
		\alpha^{-2}\epsilon^{-k_x} & 0 & \alpha^{2}\epsilon^{-k_y} & \alpha\,\epsilon^{k_x} & \epsilon^{k_x}\epsilon^{k_y} & \alpha^{-1}\epsilon^{k_y}\\
		\alpha^{-1}\epsilon^{-k_x} & \alpha^{-2}\epsilon^{-k_x}\epsilon^{-k_y} & 0 & \alpha^{2}\epsilon^{k_x} & \alpha\,\epsilon^{k_x}\epsilon^{k_y} & \epsilon^{k_y}
	\end{pmatrix}
	\begin{pmatrix}
		\tilde{W}_{r}(k_x,k_y,1)\\
		\tilde{W}_{r}(k_x,k_y,2)\\
		\tilde{W}_{r}(k_x,k_y,3)\\
		\tilde{W}_{r}(k_x,k_y,4)\\
		\tilde{W}_{r}(k_x,k_y,5)\\
		\tilde{W}_{r}(k_x,k_y,6)
	\end{pmatrix}, 
    \label{FTreccurenceRelation}
\end{align}
where $\alpha = e^{\,i\pi/6}$
and $\epsilon = e^{\,\frac{2\pi i}{L}}$.
The partition function of the system can be expressed, using Eq.~\eqref{R_1} through the eigenvalues of the square $6\times 6$ matrix in Eq.~\eqref{FTreccurenceRelation}:

\begin{align}
	R &= 2^{N_s} \prod_{k_x,k_y=0}^{L} \prod_{i=1}^{6}\left( 1 - t\,\tilde{\lambda}_{i} \right)^{1/2}
	= 2^{N_s} \prod_{k_x,k_y=0}^{L} \det\left(\delta_{\nu\nu'} - t\,\tilde{\Lambda}_{\nu\nu'}\right)^{1/2} \nonumber\\
	&= 2^{N_s} \prod_{k_x,k_y=0}^{L} \left\{ 1 + t^2 \left[ 3 + t \left(8 + 3t + t^3 \right) \right] 
	- 2t \left(1 - t^2 \right)^2 \left[ \cos\left(\frac{2\pi k_x}{L}\right) + \cos\left(\frac{2\pi k_y}{L}\right) + \cos\left(\frac{2\pi k_x+ 2\pi k_y}{L}\right) \right] \right\}^{1/2} \nonumber\\
	&=  2^{N_s}  \left(1+t\right)^{N_s}  \prod_{k_x,k_y=0}^{L} \left\{ \left(t^4 - 2t^3 + 6t^2 - 2t + 1 \right) - 2t \left(1-t \right)^2 \left[ \cos\left(\frac{2\pi k_x}{L}\right) + \cos\left(\frac{2\pi k_y}{L}\right) + \cos\left(\frac{2\pi k_x+ 2\pi k_y}{L}\right) \right] \right\}^{1/2}. \label{eq:R_prod}
\end{align}

Finally, we obtain the partition function for the triangular system as
\begin{align}
	Z_0 &= 2^{N_s} (1 - t^2)^{-\frac{3N_s}{2}} \left(1+t\right)^{N_s}  \times \nonumber\\
	& \quad \prod_{k_x,k_y=0}^{L} \left\{ \left(t^4 - 2t^3 + 6t^2 - 2t + 1 \right) - 2t \left(1-t \right)^2 \left[ \cos\left(\frac{2\pi k_x}{L}\right) + \cos\left(\frac{2\pi k_y}{L}\right) + \cos\left(\frac{2\pi k_x+ 2\pi k_y}{L}\right) \right] \right\}^{1/2}, 
    \label{PartitionFunction0}
\end{align}
where we used that $N_b = 3 N_s$ for the triangular lattice.
The free energy of the system can be obtained from the partition
function~\eqref{PartitionFunction0}, which gives
\begin{align}
    F_0 = & -T \ln Z \nonumber\\
    = & -N_sT\ln 2 + \frac{3N_s}{2}T \ln (1 - t^2) - N_sT\ln (1 + t) \nonumber\\
    &- \frac{1}{2} T \sum_{k_x,k_y=0}^{L} 
    \ln \left\{ \left(t^4 - 2t^3 + 6t^2 - 2t + 1 \right) - 2t \left(1-t \right)^2 \left[ \cos\left(\frac{2\pi k_x}{L}\right) + \cos\left(\frac{2\pi k_y}{L}\right) + \cos\left(\frac{2\pi k_x+ 2\pi k_y}{L}\right) \right] \right\}. \label{F0sum}
\end{align}
In the thermodynamic limit ($L \rightarrow \infty$), the sum
in Eq.~\eqref{F0sum} can be replaced by an integral:
\begin{align}
	F_0  = -N_sT\ln 2  + \frac{3N_s}{2}T \ln (1 - t^2) -N_sT\ln (1 + t) -\frac{N_sT}{2 (2\pi)^2} \int_0^{2\pi} \int_0^{2\pi} 
	\ln P \, d\omega_1 \, d\omega_2, 
    \label{freeEn}
\end{align}
where $P =  \left(t^4 - 2t^3 + 6t^2 - 2t + 1 \right) - 2t \bigl(1-t \bigr)^2 \omega$
and $ \omega = \left[\cos\omega_1 + \cos\omega_2 + \cos\left(\omega_1 + \omega_2\right) \right]$.

The free energy can be expanded in the small exponential $e^{-\frac{2J}{T}}$ as
\begin{align}
	F_0  =& N_sT\ln 2  - N_s J -2N_s Te^{-\frac{2J}{T}} + N_s Te^{-\frac{4J}{T}}
	\nonumber \\
		& + 2N_s Te^{-\frac{2J}{T}} -N_s T \ln2 -\frac{N_s T}{8 \pi^2} \int_0^{2\pi} \int_0^{2\pi} \left[ \ln\left(3 + 2 \omega\right) + 2\frac{3+\omega}{3+2\omega} e^{-\frac{4J}{T}} \right] \, d\omega_1 \, d\omega_2 + O\left(e^{-\frac{6J}{T}} \right)
	\nonumber \\
	= & - N_s J - N_sT I_S + \left(1 - I_0\right) N_s T e^{-\frac{4J}{T}}
	 + O\left(e^{-\frac{6J}{T}} \right),
\end{align}
where we have defined 
\begin{align}
	I_0  & \equiv  \frac{1}{4 \pi^2} \int_0^{2\pi} \int_0^{2\pi} 
    \frac{3+\omega}{3+2\omega}  \, d\omega_1 \, d\omega_2
	=  5.191389\ldots,
    \label{I0}
\end{align}
 and 
\begin{align}
	I_S  \equiv & \frac{1}{8 \pi^2} \int_0^{2\pi} \int_0^{2\pi} 
     \ln\left(3 + 2 \omega\right)   \, d\omega_1 \, d\omega_2 
	=  0.323066\ldots.
\end{align}
In what follows, we will see that $I_S$ gives the entropy of the system per spin at zero
temperature.

Utilising Eq.~\eqref{freeEn}, we obtain the entropy of the system per spin:
\begin{align}
	S_0(T) = & -\frac{1}{N_s}\frac{\partial F_0}{\partial T} \nonumber\\
	= & \ln 2 -\frac{3}{2}\left[\ln\left(1-t^2\right) - \frac{2Jt}{T}\right] +\left[\ln(1+t) + \frac{\left(1-t\right)J}{T}\right]
	+ \frac{1}{8\pi^2} \int_0^{2\pi} \int_0^{2\pi} \left( \ln P + T\,\frac{1}{P}\frac{\partial P}{\partial t}\frac{dt}{dT} \right) d\omega_1 \, d\omega_2. \label{entropy}
\end{align}


Expanding the above entropy $S_0(T)$ in the small exponential $e^{-\frac{2J}{T}}$ gives
\begin{align}
	S_0(T) = & -\ln2 + \frac{1}{8\pi^2}\int_0^{2\pi}\int_0^{2\pi}\ln\left(12+8\omega\right)d\omega_1\,d\omega_2
	\nonumber \\
	&+\left[ \left( 2 + \frac{4J}{T} \right)
	-\frac{1}{8\pi^2}\int_0^{2\pi}\int_0^{2\pi}\left( 4 + \frac{8J}{T} \right)d\omega_1\,d\omega_2\right]e^{-\frac{2J}{T}}
	\nonumber \\
	&+ \left( 1 + \frac{4J}{T} \right) e^{-\frac{4J}{T}} \left(\frac{1}{4\pi^2}\int_0^{2\pi}\int_0^{2\pi} \frac{3 + \omega}{3 + 2\omega}d\omega_1\,d\omega_2  -1\right)
	+O\left(e^{-\frac{6J}{T}} \right) 
	\nonumber \\
	 = & I_S + \left(I_0 - 1\right) \left( 1 + \frac{4J}{T} \right)  e^{-\frac{4J}{T}} + O\left(e^{-\frac{6J}{T}} \right). 
     \label{entropy_1}
\end{align}

The zero-point entropy is given by
\begin{align}
	S_0(0) &= I_S = \frac{1}{8\pi^2} \int_0^{2\pi} \int_0^{2\pi} \ln \left\{3 + 2\left[\cos\omega_1 + \cos\omega_2 + \cos\left(\omega_1 + \omega_2\right)\right]\right\} \, d\omega_1 \, d\omega_2 \nonumber\\
	&= \frac{1}{8\pi^2} \int_0^{2\pi} \int_0^{2\pi} \ln \left[ 1 + 4 \cos\left(\frac{\omega_1 + \omega_2}{2}\right)\cos\left(\frac{\omega_1 - \omega_2}{2}\right) + 4\cos^2\left(\frac{\omega_1 + \omega_2}{2}\right) \right] \, d\omega_1 \, d\omega_2 \nonumber\\
	&= \frac{1}{8\pi^2} \int_0^{2\pi} \int_0^{2\pi} \ln \left( 1 - 4 \cos \omega \cos\omega' + 4\cos^2\omega' \right) \, d\omega \, d\omega' \nonumber\\
	&\approx  0.323066\ldots. 
    \label{ZeroTemperatureEntropyNoVacancy}
\end{align}
The value of the entropy~\eqref{ZeroTemperatureEntropyNoVacancy}
matches the result obtained by Wannier~\cite{Wannier:Ising,Wannier:erratum}
using a different method.

The heat capacity per site is given by 
\begin{align}
	C_0(T) = T\partial_TS_0(T) &= 3\left(\frac{T\,t}{1-t^2}\frac{dt}{dT}+J\,\frac{dt}{dT}-\frac{J\,t}{T}\right)
	+\frac{T}{1+t}\frac{dt}{dT}-J\,\frac{dt}{dT}-\frac{J(1-t)}{T}
	\nonumber \\
	&\quad+\frac{T}{8\pi^2}\int_0^{2\pi}\int_0^{2\pi}\left\{\frac{2}{P}\frac{\partial P}{\partial t}\frac{dt}{dT}
	+ T\left[\frac{P\,\frac{\partial^2 P}{\partial t^2}-\left(\frac{\partial P}{\partial t}\right)^2}{P^2}\left(\frac{dt}{dT}\right)^2
	+\frac{1}{P}\frac{\partial P}{\partial t}\frac{d^2t}{dT^2}\right]\right\}d\omega_1\,d\omega_2\,. 
\end{align}

Expanding the heat capacity $C_0(T)$ in the small parameter $e^{-\frac{2J}{T}}$
gives
\begin{align}
	C_0(T) =& \frac{8J^2}{T^2}\,e^{-2\frac{J}{T}} -\frac{16J^2}{T^2}e^{-\frac{4J}{T}}
         +\frac{1}{8\pi^2}\int_0^{2\pi}\int_0^{2\pi} \left[-\frac{16 J^2 }{T^2} e^{-\frac{2J}{T}} + \frac{32 J^2\left( 3 + \omega \right) }{T^2 (3 + 2\omega)} e^{-\frac{4J}{T}} \right] d\omega_1\,d\omega_2\, + O \left( e^{-\frac{6J}{T}} \right)
         \nonumber \\
         =& \frac{16J^2}{T^2} (I_0 - 1) e^{-\frac{4J}{T}}  + O \left(e^{-\frac{6J}{T}}  \right),
         \label{CleanCAppendix}
\end{align}
where the constant $I_0$ is defined by Eq.~\eqref{I0}.


\subsection*{Activation gap}

The heat capacity~\eqref{CleanCAppendix}  
shows exponential behavior with the activation gap $4J$, which corresponds to the minimum excitation energy in the Ising model on the triangular lattice.
The value of the gap can be understood as follows.
Any configuration of spin can be obtained from any other configuration by consecutively flipping individual spins. For each such individual flip,
the energy of the system changes by a multiple of $4J$.
Indeed, if a spin has $n$ ``down'' neighbours and $6-n$ ``up'' neighbours
and flips from ``down'' to ``up'', the change of the energy is given by 
$4J(n-3)$, which is a multiple of $4J$.
It can straightforwardly be verified that excited states with the energy $4J$,
measured from the ground-state energy,
exist in the system. The respective gap reflects in the heat capacity~\eqref{CleanCAppendix}, free energy~\eqref{freeEn}
and entropy~\eqref{entropy_1}. Let us demonstrate that 
no excitations with smaller energies contribute to the 
free energy.

\begin{figure}[h]
    \centering
    \includegraphics[width=0.5\linewidth]{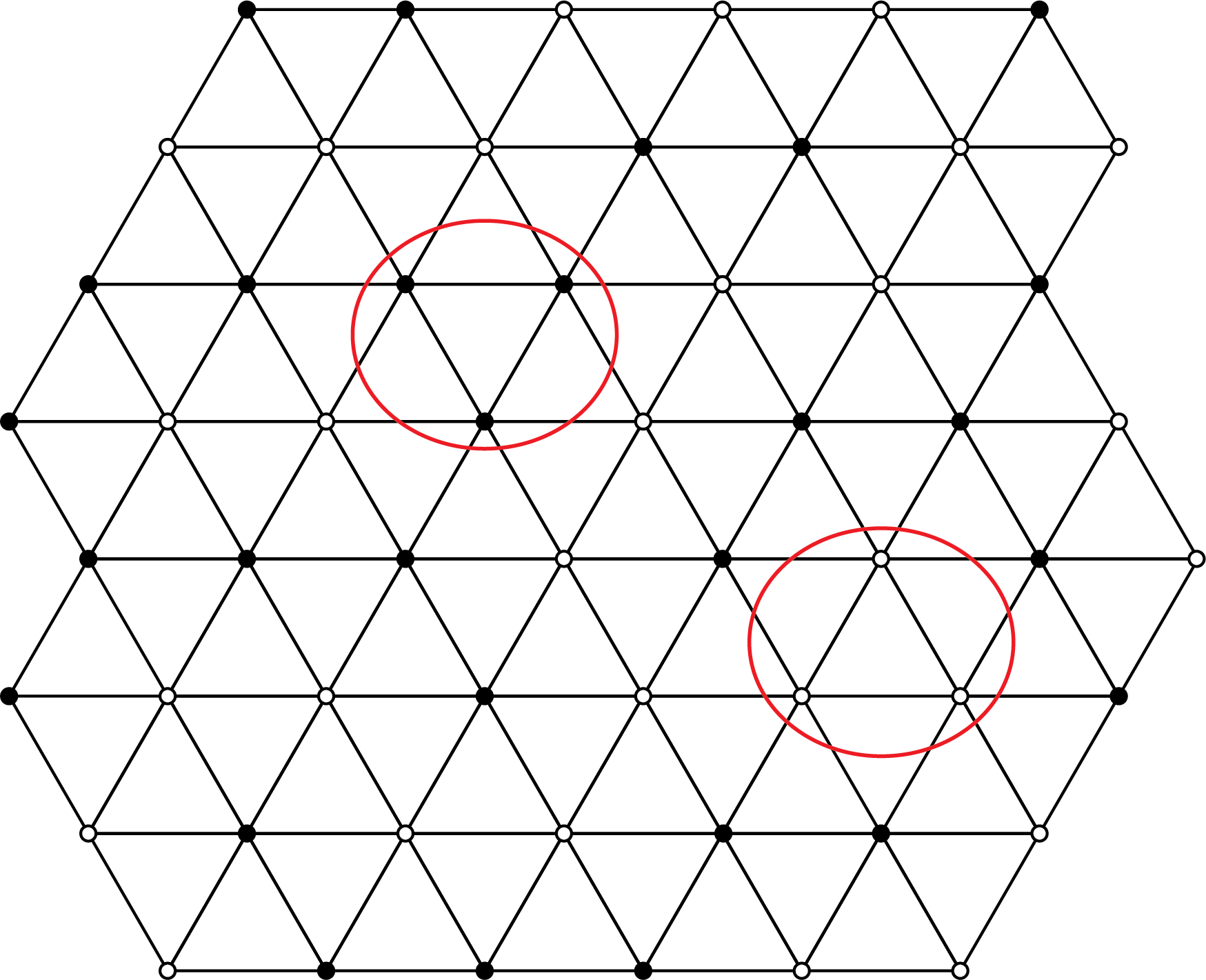}
    \caption{The state of the 
    antiferromagnetic Ising model on the triangular lattice with two ``fractionalized'' excitations. A black (open) circle 
    at a node represents the ``up'' (``down'') state at this node.
    A ``fractionalized'' excitation has an energy of $2J$ and corresponds to a triangle of three unsatisfied bonds. The other
    triangles, where the excitations are absent, have one satisfied and two unsatisfied bonds.}
    \label{fig:vortices}
\end{figure}

The triangular lattice allows for ``fractionalized'' excitations~\cite{Landau:Vortices} of energy $2J$, shown in Fig.~\ref{fig:vortices}.
Such excitations represent triangles of three unsatisfied bonds,
among the other triangles that have one unsatisfied and two satisfied bonds, as in the Ising ground states. 

Although the energy $2J$ of a fractionalized excitation is finite,
its free energy is logarithmically divergent~\cite{Moore:LogInteractions,Chalker:notes},
similar to the energy of vortices in the XY model, which is why vortices
appear in the spin configuration as effectively bound vortex-antivortex pairs. 
The energy $4J$ of a spin flip or a vortex-antivortex pair gives the 
leading activation gap in Eq.~\eqref{CleanCAppendix}.


\section{Thermodynamics of the system with a single vacancy}


In this section, we compute the contribution of a single vacancy defect to the partition function following the approach described in Sec.~\ref{SuppSec:vacancy-free}. We will denote the single-vacancy contribution to different quantities of the system with the subscript $1$. Assuming there's a vacancy at site $\brho_0$ and using Eqs.~(\ref{eq:Z_in_R}-\ref{eq:R_0}), 
the partition function of the Ising spins on an arbitrary lattice with a single vacancy defect can be 
written as
\begin{align}
\label{eq:Z1_all}
    Z=2^{N_s-1}\left(1-t^2\right)^{-(N_b-z)/2}\exp\left(-\sum_{r=1}^{\infty}t^rf'_{r}\right),
\end{align}
where $f'_{r}$ is the sum over all loops of length $r$ that do not include the site $\brho_0$,
and the variable $t$ is defined by Eq.~\eqref{tDefinition}.

Equation~\eqref{eq:Z1_all} reflects that the vacancy 
reduces the number of sites by $1$ and the number of bonds 
by the coordination number $z$.
The sum $f'_{r}$ is given by the difference $f'_{r}=f_r-\tilde{f}_r(\brho_0)$
of the sum $f_r$ over all loops of length $r$ and the sum $\tilde{f}_r(\brho_0)$ over the loops
of length $r$ containing the site $\brho_0$. Averaging over the location of the vacancy gives 
\begin{align}
    \frac{1}{N_s}\sum_{\brho_0}\tilde{f}_r(\brho_0)=\frac{r}{N_s}f_r,
\end{align}
where we used that $\sum_{\brho_0}\tilde{f}_r(\brho_0) = r f_r$,
considering that each loop that contributes to the sum has $r$ sites.
Utilising that
\begin{align}
        f'_{r}=f_r-\frac{r}{N_s}f_r,
\end{align}
the partition function of the system with a vacancy defect can be rewritten in the form
\begin{align}
    \label{eq:Z1_Zvac}
    Z=Z_0\left[\frac{1}{2}\left(1-t^2\right)^{z/2}\exp\left(\frac{1}{N_s}\sum_{r=1}^{\infty}rt^rf_r\right)\right]\equiv Z_0 Z_1.
\end{align}

The sum of the series in Eq.~\eqref{eq:Z1_Zvac} can be reduced to the 
previously computed sum by differentiating with respect to $t$:
$\sum_{r=1}^{\infty}rt^rf_r=t\partial_t\left(\sum_{r=1}^{\infty}t^rf_r\right)$, which gives
\begin{align}
     \label{eq:Zvac}
    Z_1=\frac{1}{2}\left(1-t^2\right)^{z/2}\exp\left[\frac{t}{N_s}\frac{\partial}{\partial t}\left(\sum_{r=1}^{\infty}t^rf_r\right)\right].
\end{align}
Equation~\eqref{eq:Zvac} describes the partition function of a vacancy
defect in a generic lattice. In what follows, we explicitly compute 
this function for a triangular lattice.

The sum $\sum_{r=1}^{\infty}t^rf_r$ has been computed in 
Sec.~\ref{SuppSec:vacancy-free} and is given by Eqs.~\eqref{eq:R_0}
and \eqref{eq:R_prod}:
\begin{align}
    &\sum_{r=1}^{\infty}t^rf_r=
    \nonumber\\
    &-N_s \ln(1+t)-\frac{1}{2}\sum_{k_x,k_y=0}^{L}\ln 
	\left\{ \left(t^4 - 2t^3 + 6t^2 - 2t + 1 \right) - 2t \left(1-t \right)^2 \left[ \cos\left(\frac{2\pi k_x}{L}\right) + \cos\left(\frac{2\pi k_y}{L}\right) + \cos\left(\frac{2\pi k_x+ 2\pi k_y}{L}\right) \right] \right\}.
    \label{SumForVacancy}
\end{align}

Utilising Eqs.~\eqref{eq:Zvac} and \eqref{SumForVacancy}, we arrive at the free energy
$F_1=F-F_0$ associated with the vacancy defect:
\begin{align}
    F_1=T\ln(2)-3T\ln(1-t^2)+\frac{Tt}{1+t}+\frac{T}{2(2\pi)^2}\int_0^{2\pi}\int_0^{2\pi}\frac{t\partial_tP(t,\omega_1,\omega_2)}{P(t,\omega_1,\omega_2)}\mathrm{d}\omega_1\mathrm{d}\omega_2. \label{Eq:Fvac}
\end{align}

In the limit of low temperatures $T\ll J$, the variable $t$ can be approximated as
$t\equiv -\tanh\left(J/T\right)\approx (-1)\left(1-2e^{-2J/T}+2e^{-4J/T}\right)+O(e^{-6J/T})$
and the free energy $F_1$ of the vacancy defect can be expanded in the 
powers of the small exponential $e^{-\frac{2J}{T}}\ll 1$ as
\begin{align}
    F_1= -\frac{1}{2}Te^{\frac{2J}{T}}+6J-T\left[5\ln(2)-\frac{3}{2}\right]+\left(6-I_0\right)Te^{-\frac{2J}{T}}+3Te^{-\frac{4J}{T}}+O\left(e^{-\frac{6J}{T}}\right).
    \label{FreeEnergyExpanded}
\end{align}
The leading temperature dependence in Eq.~\eqref{FreeEnergyExpanded}, given by the first term, shows exponential 
dependence of the free energy associated with the vacancy defect, on temperature. Qualitatively, this dependence is explained in the main text.

The contribution $S_1=S-S_0=-\partial_T F_1$ of the vacancy defect to the entropy is given by
\begin{align}
    S_1=&-\ln(2)+3\left[\ln(1-t^2)-\frac{2Jt}{T}\right]-\frac{t}{1+t}-\frac{J(1-t)}{T(1+t)}\notag\\
    & -\frac{1}{2}\int_0^{2\pi}\int_0^{2\pi}\left\{\frac{t\partial_tP(t,\omega_1,\omega_2)}{P(t,\omega_1,\omega_2)}+\frac{J(1-t^2)}{T}\frac{\partial}{\partial t}\left[\frac{t\partial_tP(t,\omega_1,\omega_2)}{P(t,\omega_1,\omega_2)}\right]\right\}\mathrm{d}\omega_1\mathrm{d}\omega_2.
\end{align}
In the limit of low temperatures $T\ll J$,
the entropy associated with the vacancy is given by
\begin{align}
     S_1= \frac{1}{2}\left(1-\frac{2J}{T}\right)e^{\frac{2J}{T}}+5\ln(2)-\frac{3}{2}-\left(6-I_0\right)\left(1+\frac{2J}{T}\right)e^{-\frac{2J}{T}}+3\left(1+\frac{4J}{T}\right) e^{-\frac{4J}{T}}+O\left(e^{-\frac{6J}{T}}\right).
     \label{eqn:Svac_low_T}
\end{align}
The vacancy contribution $C_1=C-C_0$ to the heat capacity is given by
\begin{align}
      C_1=&T\partial_TS_{1}(T)
      =\frac{J^2}{T^2}(1-t^2)\left[-6+\frac{2}{(1+t)^2}\right]\notag\\
      +&\frac{J^2}{T^2}(1-t^2)\int_0^{2\pi}\int_0^{2\pi}\left\{t \partial_t \ln[P(t,\omega_1,\omega_2)]-(1-2t^2)\partial^2_t\ln[P(t,\omega_1,\omega_2)]-\frac{1}{2}(1-t^2)\partial^2_t\ln[P(t,\omega_1,\omega_2)]\right\}\mathrm{d}\omega_1\mathrm{d}\omega_2.
\end{align}
In the limit of low temperatures $T\ll J$, the heat capacity associated with the vacancy is given by 
\begin{align}
      C_1=\frac{2J^2}{T^2}\left[e^{\frac{2J}{T}}-2\left(6-I_0\right)e^{-\frac{2J}{T}}+24e^{-\frac{4J}{T}}\right]+O\left(e^{-\frac{6J}{T}}\right).
\end{align}


\section{Thermodynamics of the system with dilute vacancies}

In a system with dilute vacancy defects density $n_{\text{imp}}=N_{\text{imp}}/N_s\ll \xi^{-2}$, where $\xi$ is the correlation length, the 
interplay of different vacancies can be neglected due to the large typical
distance between the vacancies.
In this regime, the contributions of vacancy defects
to the entropy and the free energy are additive.
At low temperatures $T\ll J$,  
\begin{align}
    \frac{F}{N_s}=\frac{1}{N_s}F_0+n_{\text{imp}} F_1&= -J- T I_S + \left(1 - I_0\right) T e^{-\frac{4J}{T}}\notag\\
    &\quad +n_{\text{imp}} \left\{6J-\frac{1}{2}Te^{\frac{2J}{T}}-T\left[5\ln(2)-\frac{3}{2}\right]+\left(6-I_0\right)Te^{-\frac{2J}{T}}+3Te^{-\frac{4J}{T}}\right\}+O\left(e^{-\frac{6J}{T}}\right).
\end{align}

The corresponding entropy is given by
\begin{align}
    S=&I_S + \left(I_0 - 1\right) \left( 1 + \frac{4J}{T} \right)  e^{-\frac{4J}{T}}\notag\\
    & +n_{\text{imp}}\left[\frac{1}{2}\left(1-\frac{2J}{T}\right)e^{\frac{2J}{T}}+5\ln(2)-\frac{3}{2}-\left(6-I_0\right)\left(1+\frac{2J}{T}\right)e^{-\frac{2J}{T}}+3\left(1+\frac{4J}{T}\right) e^{-\frac{4J}{T}}\right]+O\left(e^{-\frac{6J}{T}}\right).
\end{align}
The vacancies contribution is $S_\text{vac}=N_\text{imp} S_1$ in this dilute vacancy density limit. Finally, the heat capacity of the system with dilute vacancy defects
is given by 
\begin{align}
    C=\frac{2J^2}{T^2}\left\{8 (I_0 - 1) e^{-\frac{4J}{T}}+n_{\text{imp}}\left[e^{\frac{2J}{T}} -2\left(6-I_0\right)e^{-\frac{2J}{T}} +24e^{-\frac{4J}{T}}\right]\right\}+ O\left(e^{-\frac{6J}{T}}  \right).
    \label{FullCLong}
\end{align}
The term $\propto e^{-\frac{4J}{T}}$ in Eq.~\eqref{FullCLong} describes the thermodynamics of the clean system. The activation gap $4J$ 
of the exponential temperature dependence matches the lowest excitation energy in such a system, as discussed in Sec.~\ref{SuppSec:vacancy-free}.
The exponentially growing term $\propto n_\text{imp}e^\frac{2J}{T}$
describes the contribution of the vacancies to the heat capacity.

\section{Heat Capacity Structure of GF Quantum Systems}
The additive contribution of dilute vacancies to the entropy and heat capacity is comparable to the heat capacity of clean system in a temperature scale set by $T_\text{imp}\sim -J/\ln(n_\text{imp})$. This leads to a peak in the heat capacity of the system with impurities. 

\begin{figure}[h]
    \centering \includegraphics[width=0.95\linewidth]{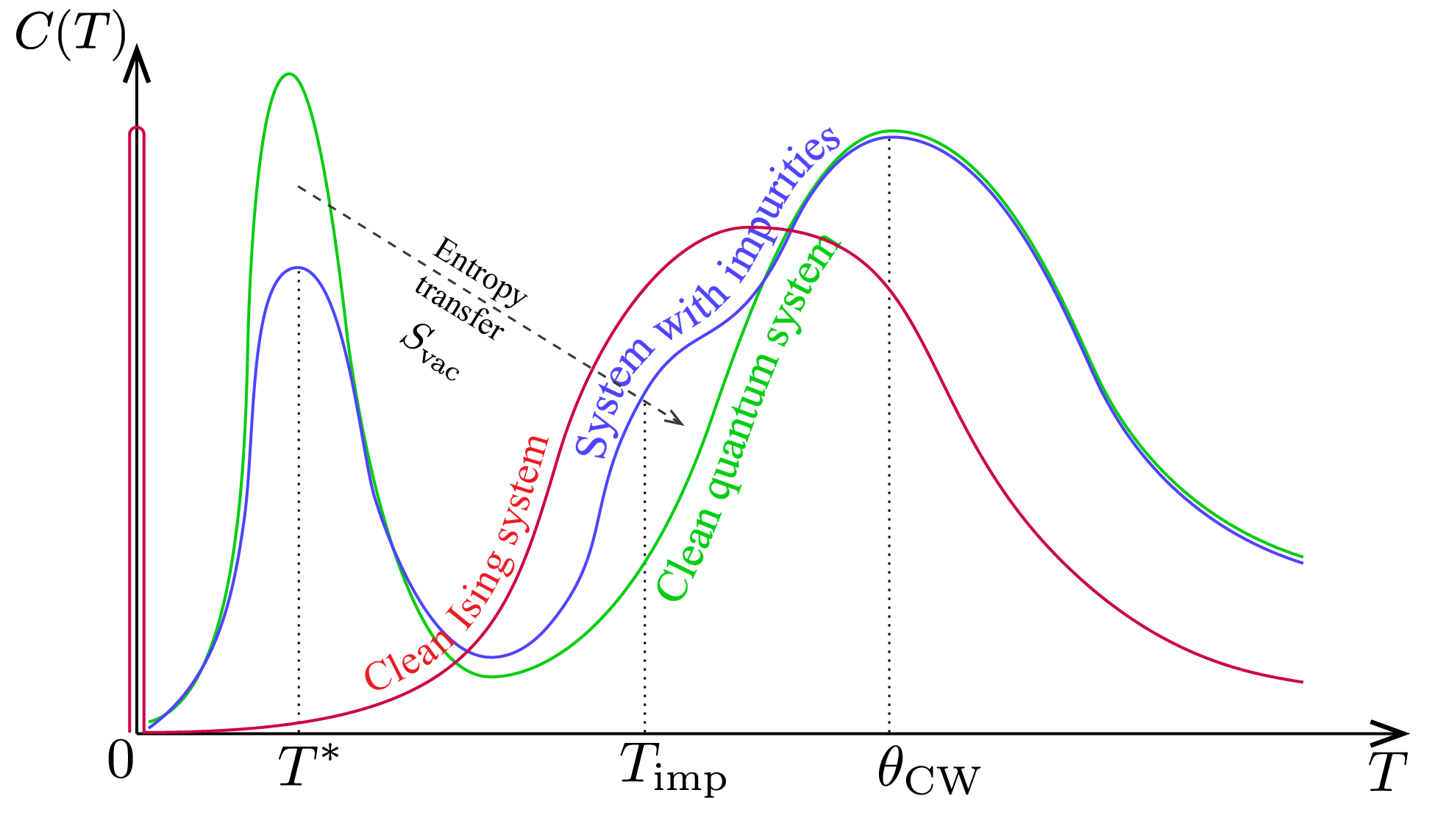}
    \caption{\label{fig:HeatCapacityWithIsing} The heat capacity of a 2D frustrated magnet in a quantum impurity-free system (green line) and 
    in a quantum system with vacancy impurities (blue line).
    Vacancy defects lead to an anomaly that manifests itself as a peak at a temperature
    $T_\text{imp} \sim -J/\ln(n_\text{imp})$, which is in general
    distinct from the other temperature scales in
    the system. In a generic frustrated system, the illustrated 
    heat-capacity peaks may partially merge. The peak at $T=0$ of the clean Ising system capacity is only illustrative and is meant to represent limit of the $T^*$-peak in the Heisenberg model as the transverse coupling vanishes. 
    }
\end{figure}

This vacancy-induced peak in the heat capacity persists also in GF Heisenberg magnets. The clean systems of such magnets are known to exhibit two-peak structure of the heat capacity: A Curie-Weiss peak at $T\sim \theta_\text{CW}$ and a low-temperature peak at $T\sim T^*$. The latter comes from the transverse coupling in the Heisenberg model hybridizing the Ising ground states \cite{PoppRamirezSyzranov,RamirezSyzranov:review}. This two peak-structure in GF quantum systems has been found numerically for triangular lattice \cite{PrelovsekKokalj:THAF,ChenQu:THAF,SekiYunoki:RingExchange,UlagaPrelovsek:latest}, Kagome lattice \cite{Elser:KHAF,SugiuraShimizu:KHAF,shimokawa2016finite,SchnackSchulenberg:KHAF}, and experimentally \cite{Li:S2TAF,Bordelon:NaYbO2,Ranjith:NaYbSe2}. A schematic illustration of the different peaks of a clean quantum system, a clean Ising system and a system with vacancies is shown in Fig. \ref{fig:HeatCapacityWithIsing}.

Due to the vacancy-induced peak, the entropy released by a material in the corresponding temperature interval increases $
    S(T)=\int_{\text{vacancy peak}}\frac{C(T)}{T}dT$. The magnitude of this entropy is 
    \begin{align}
        \left|S_\text{vac}(T_\text{imp})\right|=N_\text{imp} \left|S_1(T_\text{imp})\right|\sim N_\text{imp}\left|-\frac{J}{T_\text{imp}}e^{\frac{2J}{T_\text{imp}}}\right|\approx N_\text{imp} n_\text{imp}^{-\frac{1}{2}}=(N_s N_\text{imp})^{\frac{1}{2}},
    \end{align}
    where we use $T_\text{imp}=-4J/\ln(n_\text{imp})$. This entropy comes from the Ising system ground state as the vacancies lift the ground state degeneracy. In a quantum magnet, it comes from the lower temperature peak at $T\sim T^*$.

\putbib[references]
\end{bibunit}


\begin{thebibliography}{42}%
\makeatletter
\providecommand \@ifxundefined [1]{%
 \@ifx{#1\undefined}
}%
\providecommand \@ifnum [1]{%
 \ifnum #1\expandafter \@firstoftwo
 \else \expandafter \@secondoftwo
 \fi
}%
\providecommand \@ifx [1]{%
 \ifx #1\expandafter \@firstoftwo
 \else \expandafter \@secondoftwo
 \fi
}%
\providecommand \natexlab [1]{#1}%
\providecommand \enquote  [1]{``#1''}%
\providecommand \bibnamefont  [1]{#1}%
\providecommand \bibfnamefont [1]{#1}%
\providecommand \citenamefont [1]{#1}%
\providecommand \href@noop [0]{\@secondoftwo}%
\providecommand \href [0]{\begingroup \@sanitize@url \@href}%
\providecommand \@href[1]{\@@startlink{#1}\@@href}%
\providecommand \@@href[1]{\endgroup#1\@@endlink}%
\providecommand \@sanitize@url [0]{\catcode `\\12\catcode `\$12\catcode `\&12\catcode `\#12\catcode `\^12\catcode `\_12\catcode `\%12\relax}%
\providecommand \@@startlink[1]{}%
\providecommand \@@endlink[0]{}%
\providecommand \url  [0]{\begingroup\@sanitize@url \@url }%
\providecommand \@url [1]{\endgroup\@href {#1}{\urlprefix }}%
\providecommand \urlprefix  [0]{URL }%
\providecommand \Eprint [0]{\href }%
\providecommand \doibase [0]{https://doi.org/}%
\providecommand \selectlanguage [0]{\@gobble}%
\providecommand \bibinfo  [0]{\@secondoftwo}%
\providecommand \bibfield  [0]{\@secondoftwo}%
\providecommand \translation [1]{[#1]}%
\providecommand \BibitemOpen [0]{}%
\providecommand \bibitemStop [0]{}%
\providecommand \bibitemNoStop [0]{.\EOS\space}%
\providecommand \EOS [0]{\spacefactor3000\relax}%
\providecommand \BibitemShut  [1]{\csname bibitem#1\endcsname}%
\let\auto@bib@innerbib\@empty
\bibitem [{\citenamefont {Binder}\ and\ \citenamefont {Young}(1986)}]{BinderYoung:review}%
  \BibitemOpen
  \bibfield  {author} {\bibinfo {author} {\bibfnamefont {K.}~\bibnamefont {Binder}}\ and\ \bibinfo {author} {\bibfnamefont {A.~P.}\ \bibnamefont {Young}},\ }\href {https://doi.org/10.1103/RevModPhys.58.801} {\bibfield  {journal} {\bibinfo  {journal} {Rev. Mod. Phys.}\ }\textbf {\bibinfo {volume} {58}},\ \bibinfo {pages} {801} (\bibinfo {year} {1986})}\BibitemShut {NoStop}%
\bibitem [{\citenamefont {Savary}\ and\ \citenamefont {Balents}(2016)}]{SavaryBalents:review}%
  \BibitemOpen
  \bibfield  {author} {\bibinfo {author} {\bibfnamefont {L.}~\bibnamefont {Savary}}\ and\ \bibinfo {author} {\bibfnamefont {L.}~\bibnamefont {Balents}},\ }\href@noop {} {\bibfield  {journal} {\bibinfo  {journal} {Reports on Progress in Physics}\ }\textbf {\bibinfo {volume} {80}},\ \bibinfo {pages} {016502} (\bibinfo {year} {2016})}\BibitemShut {NoStop}%
\bibitem [{\citenamefont {Schiffer}\ and\ \citenamefont {Daruka}(1997)}]{Schiffer:TwoPopulationModel}%
  \BibitemOpen
  \bibfield  {author} {\bibinfo {author} {\bibfnamefont {P.}~\bibnamefont {Schiffer}}\ and\ \bibinfo {author} {\bibfnamefont {I.}~\bibnamefont {Daruka}},\ }\href {https://doi.org/10.1103/PhysRevB.56.13712} {\bibfield  {journal} {\bibinfo  {journal} {Phys. Rev. B}\ }\textbf {\bibinfo {volume} {56}},\ \bibinfo {pages} {13712} (\bibinfo {year} {1997})}\BibitemShut {NoStop}%
\bibitem [{\citenamefont {LaForge}\ \emph {et~al.}(2013)\citenamefont {LaForge}, \citenamefont {Pulido}, \citenamefont {Cava}, \citenamefont {Chan},\ and\ \citenamefont {Ramirez}}]{LaForge:quasispin}%
  \BibitemOpen
  \bibfield  {author} {\bibinfo {author} {\bibfnamefont {A.~D.}\ \bibnamefont {LaForge}}, \bibinfo {author} {\bibfnamefont {S.~H.}\ \bibnamefont {Pulido}}, \bibinfo {author} {\bibfnamefont {R.~J.}\ \bibnamefont {Cava}}, \bibinfo {author} {\bibfnamefont {B.~C.}\ \bibnamefont {Chan}},\ and\ \bibinfo {author} {\bibfnamefont {A.~P.}\ \bibnamefont {Ramirez}},\ }\href {https://doi.org/10.1103/PhysRevLett.110.017203} {\bibfield  {journal} {\bibinfo  {journal} {Phys. Rev. Lett.}\ }\textbf {\bibinfo {volume} {110}},\ \bibinfo {pages} {017203} (\bibinfo {year} {2013})}\BibitemShut {NoStop}%
\bibitem [{\citenamefont {Syzranov}\ and\ \citenamefont {Ramirez}(2022)}]{Syzranov:HiddenEnergy}%
  \BibitemOpen
  \bibfield  {author} {\bibinfo {author} {\bibfnamefont {S.~V.}\ \bibnamefont {Syzranov}}\ and\ \bibinfo {author} {\bibfnamefont {A.~P.}\ \bibnamefont {Ramirez}},\ }\href {https://www.nature.com/articles/s41467-022-30739-0} {\bibfield  {journal} {\bibinfo  {journal} {Nature Communications}\ }\textbf {\bibinfo {volume} {13}},\ \bibinfo {pages} {2993} (\bibinfo {year} {2022})}\BibitemShut {NoStop}%
\bibitem [{\citenamefont {Ramirez}\ and\ \citenamefont {Syzranov}(2025)}]{RamirezSyzranov:review}%
  \BibitemOpen
  \bibfield  {author} {\bibinfo {author} {\bibfnamefont {A.~P.}\ \bibnamefont {Ramirez}}\ and\ \bibinfo {author} {\bibfnamefont {S.~V.}\ \bibnamefont {Syzranov}},\ }\href {https://doi.org/10.1039/D4MA00914B} {\bibfield  {journal} {\bibinfo  {journal} {Mater. Adv.}\ }\textbf {\bibinfo {volume} {6}},\ \bibinfo {pages} {1213} (\bibinfo {year} {2025})}\BibitemShut {NoStop}%
\bibitem [{\citenamefont {Elser}(1984)}]{Elser:BoundaryDependence}%
  \BibitemOpen
  \bibfield  {author} {\bibinfo {author} {\bibfnamefont {V.}~\bibnamefont {Elser}},\ }\href {https://doi.org/10.1088/0305-4470/17/7/018} {\bibfield  {journal} {\bibinfo  {journal} {Journal of Physics A: Mathematical and General}\ }\textbf {\bibinfo {volume} {17}},\ \bibinfo {pages} {1509} (\bibinfo {year} {1984})}\BibitemShut {NoStop}%
\bibitem [{\citenamefont {Ferreyra}\ and\ \citenamefont {Grigera}(2018)}]{Ferreyra:BD_one}%
  \BibitemOpen
  \bibfield  {author} {\bibinfo {author} {\bibfnamefont {M.~V.}\ \bibnamefont {Ferreyra}}\ and\ \bibinfo {author} {\bibfnamefont {S.~A.}\ \bibnamefont {Grigera}},\ }\href {https://doi.org/10.1103/PhysRevE.98.042146} {\bibfield  {journal} {\bibinfo  {journal} {Phys. Rev. E}\ }\textbf {\bibinfo {volume} {98}},\ \bibinfo {pages} {042146} (\bibinfo {year} {2018})}\BibitemShut {NoStop}%
\bibitem [{\citenamefont {Tavares}\ \emph {et~al.}(2015{\natexlab{a}})\citenamefont {Tavares}, \citenamefont {Ribeiro},\ and\ \citenamefont {Korepin}}]{Tavares:BD_two}%
  \BibitemOpen
  \bibfield  {author} {\bibinfo {author} {\bibfnamefont {T.~S.}\ \bibnamefont {Tavares}}, \bibinfo {author} {\bibfnamefont {G.~A.~P.}\ \bibnamefont {Ribeiro}},\ and\ \bibinfo {author} {\bibfnamefont {V.~E.}\ \bibnamefont {Korepin}},\ }\href {https://doi.org/10.1088/1742-5468/2015/06/P06016} {\bibfield  {journal} {\bibinfo  {journal} {Journal of Statistical Mechanics: Theory and Experiment}\ }\textbf {\bibinfo {volume} {2015}},\ \bibinfo {pages} {P06016} (\bibinfo {year} {2015}{\natexlab{a}})}\BibitemShut {NoStop}%
\bibitem [{\citenamefont {Tavares}\ \emph {et~al.}(2015{\natexlab{b}})\citenamefont {Tavares}, \citenamefont {Ribeiro},\ and\ \citenamefont {Korepin}}]{Tavares:BD_three}%
  \BibitemOpen
  \bibfield  {author} {\bibinfo {author} {\bibfnamefont {T.~S.}\ \bibnamefont {Tavares}}, \bibinfo {author} {\bibfnamefont {G.~A.~P.}\ \bibnamefont {Ribeiro}},\ and\ \bibinfo {author} {\bibfnamefont {V.~E.}\ \bibnamefont {Korepin}},\ }\href {https://doi.org/10.1088/1751-8113/48/45/454004} {\bibfield  {journal} {\bibinfo  {journal} {Journal of Physics A: Mathematical and Theoretical}\ }\textbf {\bibinfo {volume} {48}},\ \bibinfo {pages} {454004} (\bibinfo {year} {2015}{\natexlab{b}})}\BibitemShut {NoStop}%
\bibitem [{\citenamefont {McMillan}(1984)}]{McMillan:firstNoTwoDGlass}%
  \BibitemOpen
  \bibfield  {author} {\bibinfo {author} {\bibfnamefont {W.~L.}\ \bibnamefont {McMillan}},\ }\href {https://doi.org/10.1103/PhysRevB.30.476} {\bibfield  {journal} {\bibinfo  {journal} {Phys. Rev. B}\ }\textbf {\bibinfo {volume} {30}},\ \bibinfo {pages} {476} (\bibinfo {year} {1984})}\BibitemShut {NoStop}%
\bibitem [{\citenamefont {Hartmann}\ and\ \citenamefont {Young}(2001)}]{HartmannYoung:TwoDGlass}%
  \BibitemOpen
  \bibfield  {author} {\bibinfo {author} {\bibfnamefont {A.~K.}\ \bibnamefont {Hartmann}}\ and\ \bibinfo {author} {\bibfnamefont {A.~P.}\ \bibnamefont {Young}},\ }\href {https://doi.org/10.1103/PhysRevB.64.180404} {\bibfield  {journal} {\bibinfo  {journal} {Phys. Rev. B}\ }\textbf {\bibinfo {volume} {64}},\ \bibinfo {pages} {180404} (\bibinfo {year} {2001})}\BibitemShut {NoStop}%
\bibitem [{\citenamefont {Carter}\ \emph {et~al.}(2002)\citenamefont {Carter}, \citenamefont {Bray},\ and\ \citenamefont {Moore}}]{Carter:TwoDGlass}%
  \BibitemOpen
  \bibfield  {author} {\bibinfo {author} {\bibfnamefont {A.~C.}\ \bibnamefont {Carter}}, \bibinfo {author} {\bibfnamefont {A.~J.}\ \bibnamefont {Bray}},\ and\ \bibinfo {author} {\bibfnamefont {M.~A.}\ \bibnamefont {Moore}},\ }\href {https://doi.org/10.1103/PhysRevLett.88.077201} {\bibfield  {journal} {\bibinfo  {journal} {Phys. Rev. Lett.}\ }\textbf {\bibinfo {volume} {88}},\ \bibinfo {pages} {077201} (\bibinfo {year} {2002})}\BibitemShut {NoStop}%
\bibitem [{\citenamefont {Amoruso}\ \emph {et~al.}(2003)\citenamefont {Amoruso}, \citenamefont {Marinari}, \citenamefont {Martin},\ and\ \citenamefont {Pagnani}}]{Amoruso:TwoDGlass}%
  \BibitemOpen
  \bibfield  {author} {\bibinfo {author} {\bibfnamefont {C.}~\bibnamefont {Amoruso}}, \bibinfo {author} {\bibfnamefont {E.}~\bibnamefont {Marinari}}, \bibinfo {author} {\bibfnamefont {O.~C.}\ \bibnamefont {Martin}},\ and\ \bibinfo {author} {\bibfnamefont {A.}~\bibnamefont {Pagnani}},\ }\href {https://doi.org/10.1103/PhysRevLett.91.087201} {\bibfield  {journal} {\bibinfo  {journal} {Phys. Rev. Lett.}\ }\textbf {\bibinfo {volume} {91}},\ \bibinfo {pages} {087201} (\bibinfo {year} {2003})}\BibitemShut {NoStop}%
\bibitem [{\citenamefont {Fernandez}\ \emph {et~al.}(2016)\citenamefont {Fernandez}, \citenamefont {Marinari}, \citenamefont {Martin-Mayor}, \citenamefont {Parisi},\ and\ \citenamefont {Ruiz-Lorenzo}}]{FernandezParisi:Ising2Dtransition}%
  \BibitemOpen
  \bibfield  {author} {\bibinfo {author} {\bibfnamefont {L.~A.}\ \bibnamefont {Fernandez}}, \bibinfo {author} {\bibfnamefont {E.}~\bibnamefont {Marinari}}, \bibinfo {author} {\bibfnamefont {V.}~\bibnamefont {Martin-Mayor}}, \bibinfo {author} {\bibfnamefont {G.}~\bibnamefont {Parisi}},\ and\ \bibinfo {author} {\bibfnamefont {J.~J.}\ \bibnamefont {Ruiz-Lorenzo}},\ }\href {https://doi.org/10.1103/PhysRevB.94.024402} {\bibfield  {journal} {\bibinfo  {journal} {Phys. Rev. B}\ }\textbf {\bibinfo {volume} {94}},\ \bibinfo {pages} {024402} (\bibinfo {year} {2016})}\BibitemShut {NoStop}%
\bibitem [{\citenamefont {Rieger}\ \emph {et~al.}(1996)\citenamefont {Rieger}, \citenamefont {Santen}, \citenamefont {Blasum}, \citenamefont {Diehl}, \citenamefont {J{\"u}nger},\ and\ \citenamefont {Rinaldi}}]{Rieger:TwoDGlass}%
  \BibitemOpen
  \bibfield  {author} {\bibinfo {author} {\bibfnamefont {H.}~\bibnamefont {Rieger}}, \bibinfo {author} {\bibfnamefont {L.}~\bibnamefont {Santen}}, \bibinfo {author} {\bibfnamefont {U.}~\bibnamefont {Blasum}}, \bibinfo {author} {\bibfnamefont {M.}~\bibnamefont {Diehl}}, \bibinfo {author} {\bibfnamefont {M.}~\bibnamefont {J{\"u}nger}},\ and\ \bibinfo {author} {\bibfnamefont {G.}~\bibnamefont {Rinaldi}},\ }\href@noop {} {\bibfield  {journal} {\bibinfo  {journal} {Journal of Physics A: Mathematical and General}\ }\textbf {\bibinfo {volume} {29}},\ \bibinfo {pages} {3939} (\bibinfo {year} {1996})}\BibitemShut {NoStop}%
\bibitem [{\citenamefont {Popp}\ \emph {et~al.}(2025)\citenamefont {Popp}, \citenamefont {Ramirez},\ and\ \citenamefont {Syzranov}}]{PoppRamirezSyzranov}%
  \BibitemOpen
  \bibfield  {author} {\bibinfo {author} {\bibfnamefont {P.}~\bibnamefont {Popp}}, \bibinfo {author} {\bibfnamefont {A.~P.}\ \bibnamefont {Ramirez}},\ and\ \bibinfo {author} {\bibfnamefont {S.}~\bibnamefont {Syzranov}},\ }\href {https://doi.org/10.1103/PhysRevLett.134.226701} {\bibfield  {journal} {\bibinfo  {journal} {Phys. Rev. Lett.}\ }\textbf {\bibinfo {volume} {134}},\ \bibinfo {pages} {226701} (\bibinfo {year} {2025})}\BibitemShut {NoStop}%
\bibitem [{\citenamefont {Wannier}(1950)}]{Wannier:Ising}%
  \BibitemOpen
  \bibfield  {author} {\bibinfo {author} {\bibfnamefont {G.~H.}\ \bibnamefont {Wannier}},\ }\href {https://doi.org/10.1103/PhysRev.79.357} {\bibfield  {journal} {\bibinfo  {journal} {Phys. Rev.}\ }\textbf {\bibinfo {volume} {79}},\ \bibinfo {pages} {357} (\bibinfo {year} {1950})}\BibitemShut {NoStop}%
\bibitem [{\citenamefont {Wannier}(1973)}]{Wannier:erratum}%
  \BibitemOpen
  \bibfield  {author} {\bibinfo {author} {\bibfnamefont {G.~H.}\ \bibnamefont {Wannier}},\ }\href {https://doi.org/10.1103/PhysRevB.7.5017} {\bibfield  {journal} {\bibinfo  {journal} {Phys. Rev. B}\ }\textbf {\bibinfo {volume} {7}},\ \bibinfo {pages} {5017} (\bibinfo {year} {1973})}\BibitemShut {NoStop}%
\bibitem [{\citenamefont {Houtappel}(1950)}]{Houtappel}%
  \BibitemOpen
  \bibfield  {author} {\bibinfo {author} {\bibfnamefont {R.}~\bibnamefont {Houtappel}},\ }\href {https://doi.org/https://doi.org/10.1016/0031-8914(50)90130-3} {\bibfield  {journal} {\bibinfo  {journal} {Physica}\ }\textbf {\bibinfo {volume} {16}},\ \bibinfo {pages} {425} (\bibinfo {year} {1950})}\BibitemShut {NoStop}%
\bibitem [{SM()}]{SM}%
  \BibitemOpen
  \href@noop {} {}\bibinfo {note} {See Supplemental Material at [URL will be inserted by publisher] for details.}\BibitemShut {Stop}%
\bibitem [{\citenamefont {Kac}\ and\ \citenamefont {Ward}(1952)}]{kac1952combinatorial}%
  \BibitemOpen
  \bibfield  {author} {\bibinfo {author} {\bibfnamefont {M.}~\bibnamefont {Kac}}\ and\ \bibinfo {author} {\bibfnamefont {J.~C.}\ \bibnamefont {Ward}},\ }\href {https://doi.org/10.1103/PhysRev.88.1332} {\bibfield  {journal} {\bibinfo  {journal} {Phys. Rev.}\ }\textbf {\bibinfo {volume} {88}},\ \bibinfo {pages} {1332} (\bibinfo {year} {1952})}\BibitemShut {NoStop}%
\bibitem [{\citenamefont {Landau}\ and\ \citenamefont {Lifshitz}(2013)}]{Landafshitz5}%
  \BibitemOpen
  \bibfield  {author} {\bibinfo {author} {\bibfnamefont {L.~D.}\ \bibnamefont {Landau}}\ and\ \bibinfo {author} {\bibfnamefont {E.~M.}\ \bibnamefont {Lifshitz}},\ }\href@noop {} {\emph {\bibinfo {title} {Statistical Physics: Volume 5}}},\ Vol.~\bibinfo {volume} {5}\ (\bibinfo  {publisher} {Elsevier},\ \bibinfo {year} {2013})\BibitemShut {NoStop}%
\bibitem [{Foo()}]{Footnote1}%
  \BibitemOpen
  \href@noop {} {}\bibinfo {note} {Walking along the same bond after intermediate steps is, strictly speaking, allowed, however, the respective diagrams will be cancelled by similar diagrams with opposite parities~\cite{Landafshitz5}.}\BibitemShut {Stop}%
\bibitem [{\citenamefont {Jacobsen}\ and\ \citenamefont {Fogedby}(1997)}]{Jacobsen:correlationLength}%
  \BibitemOpen
  \bibfield  {author} {\bibinfo {author} {\bibfnamefont {J.~L.}\ \bibnamefont {Jacobsen}}\ and\ \bibinfo {author} {\bibfnamefont {H.~C.}\ \bibnamefont {Fogedby}},\ }\href {https://doi.org/https://doi.org/10.1016/S0378-4371(97)00323-3} {\bibfield  {journal} {\bibinfo  {journal} {Physica A: Statistical Mechanics and its Applications}\ }\textbf {\bibinfo {volume} {246}},\ \bibinfo {pages} {563} (\bibinfo {year} {1997})}\BibitemShut {NoStop}%
\bibitem [{\citenamefont {Millane}\ and\ \citenamefont {Blakeley}(2004)}]{MillaneBlakeley:boundaryDependence}%
  \BibitemOpen
  \bibfield  {author} {\bibinfo {author} {\bibfnamefont {R.~P.}\ \bibnamefont {Millane}}\ and\ \bibinfo {author} {\bibfnamefont {N.~D.}\ \bibnamefont {Blakeley}},\ }\href {https://doi.org/10.1103/PhysRevE.70.057101} {\bibfield  {journal} {\bibinfo  {journal} {Phys. Rev. E}\ }\textbf {\bibinfo {volume} {70}},\ \bibinfo {pages} {057101} (\bibinfo {year} {2004})}\BibitemShut {NoStop}%
\bibitem [{\citenamefont {Destainville}(1998)}]{Destainville:boundaryDependence}%
  \BibitemOpen
  \bibfield  {author} {\bibinfo {author} {\bibfnamefont {N.}~\bibnamefont {Destainville}},\ }\href {https://doi.org/10.1088/0305-4470/31/29/005} {\bibfield  {journal} {\bibinfo  {journal} {Journal of Physics A: Mathematical and General}\ }\textbf {\bibinfo {volume} {31}},\ \bibinfo {pages} {6123} (\bibinfo {year} {1998})}\BibitemShut {NoStop}%
\bibitem [{\citenamefont {Greywall}\ and\ \citenamefont {Busch}(1989)}]{Greywall:DoublePeakHe}%
  \BibitemOpen
  \bibfield  {author} {\bibinfo {author} {\bibfnamefont {D.~S.}\ \bibnamefont {Greywall}}\ and\ \bibinfo {author} {\bibfnamefont {P.~A.}\ \bibnamefont {Busch}},\ }\href {https://doi.org/10.1103/PhysRevLett.62.1868} {\bibfield  {journal} {\bibinfo  {journal} {Phys. Rev. Lett.}\ }\textbf {\bibinfo {volume} {62}},\ \bibinfo {pages} {1868} (\bibinfo {year} {1989})}\BibitemShut {NoStop}%
\bibitem [{\citenamefont {Schiffer}\ \emph {et~al.}(1995)\citenamefont {Schiffer}, \citenamefont {Ramirez}, \citenamefont {Huse}, \citenamefont {Gammel}, \citenamefont {Yaron}, \citenamefont {Bishop},\ and\ \citenamefont {Valentino}}]{Schiffer:GGG}%
  \BibitemOpen
  \bibfield  {author} {\bibinfo {author} {\bibfnamefont {P.}~\bibnamefont {Schiffer}}, \bibinfo {author} {\bibfnamefont {A.~P.}\ \bibnamefont {Ramirez}}, \bibinfo {author} {\bibfnamefont {D.~A.}\ \bibnamefont {Huse}}, \bibinfo {author} {\bibfnamefont {P.~L.}\ \bibnamefont {Gammel}}, \bibinfo {author} {\bibfnamefont {U.}~\bibnamefont {Yaron}}, \bibinfo {author} {\bibfnamefont {D.~J.}\ \bibnamefont {Bishop}},\ and\ \bibinfo {author} {\bibfnamefont {A.~J.}\ \bibnamefont {Valentino}},\ }\href {https://doi.org/10.1103/PhysRevLett.74.2379} {\bibfield  {journal} {\bibinfo  {journal} {Phys. Rev. Lett.}\ }\textbf {\bibinfo {volume} {74}},\ \bibinfo {pages} {2379} (\bibinfo {year} {1995})}\BibitemShut {NoStop}%
\bibitem [{\citenamefont {Ishida}\ \emph {et~al.}(1997)\citenamefont {Ishida}, \citenamefont {Morishita}, \citenamefont {Yawata},\ and\ \citenamefont {Fukuyama}}]{Ishida:HeRingExchange}%
  \BibitemOpen
  \bibfield  {author} {\bibinfo {author} {\bibfnamefont {K.}~\bibnamefont {Ishida}}, \bibinfo {author} {\bibfnamefont {M.}~\bibnamefont {Morishita}}, \bibinfo {author} {\bibfnamefont {K.}~\bibnamefont {Yawata}},\ and\ \bibinfo {author} {\bibfnamefont {H.}~\bibnamefont {Fukuyama}},\ }\href {https://doi.org/10.1103/PhysRevLett.79.3451} {\bibfield  {journal} {\bibinfo  {journal} {Phys. Rev. Lett.}\ }\textbf {\bibinfo {volume} {79}},\ \bibinfo {pages} {3451} (\bibinfo {year} {1997})}\BibitemShut {NoStop}%
\bibitem [{\citenamefont {Nakatsuji}\ \emph {et~al.}(2005)\citenamefont {Nakatsuji}, \citenamefont {Nambu}, \citenamefont {Tonomura}, \citenamefont {Sakai}, \citenamefont {Jonas}, \citenamefont {Broholm}, \citenamefont {Tsunetsugu}, \citenamefont {Qiu},\ and\ \citenamefont {Maeno}}]{NakatshujiMaeno:NaGaSneutron}%
  \BibitemOpen
  \bibfield  {author} {\bibinfo {author} {\bibfnamefont {S.}~\bibnamefont {Nakatsuji}}, \bibinfo {author} {\bibfnamefont {Y.}~\bibnamefont {Nambu}}, \bibinfo {author} {\bibfnamefont {H.}~\bibnamefont {Tonomura}}, \bibinfo {author} {\bibfnamefont {O.}~\bibnamefont {Sakai}}, \bibinfo {author} {\bibfnamefont {S.}~\bibnamefont {Jonas}}, \bibinfo {author} {\bibfnamefont {C.}~\bibnamefont {Broholm}}, \bibinfo {author} {\bibfnamefont {H.}~\bibnamefont {Tsunetsugu}}, \bibinfo {author} {\bibfnamefont {Y.}~\bibnamefont {Qiu}},\ and\ \bibinfo {author} {\bibfnamefont {Y.}~\bibnamefont {Maeno}},\ }\href {https://doi.org/10.1126/science.1114727} {\bibfield  {journal} {\bibinfo  {journal} {Science}\ }\textbf {\bibinfo {volume} {309}},\ \bibinfo {pages} {1697} (\bibinfo {year} {2005})},\ \Eprint {https://arxiv.org/abs/https://www.science.org/doi/pdf/10.1126/science.1114727} {https://www.science.org/doi/pdf/10.1126/science.1114727} \BibitemShut {NoStop}%
\bibitem [{\citenamefont {Li}\ \emph {et~al.}(2019)\citenamefont {Li}, \citenamefont {Jin}, \citenamefont {Guo}, \citenamefont {Xu}, \citenamefont {Su}, \citenamefont {Feng}, \citenamefont {Liu}, \citenamefont {Zhou}, \citenamefont {Ying}, \citenamefont {Li}, \citenamefont {Wang}, \citenamefont {Chen},\ and\ \citenamefont {Chen}}]{Li:S2TAF}%
  \BibitemOpen
  \bibfield  {author} {\bibinfo {author} {\bibfnamefont {K.}~\bibnamefont {Li}}, \bibinfo {author} {\bibfnamefont {S.}~\bibnamefont {Jin}}, \bibinfo {author} {\bibfnamefont {J.}~\bibnamefont {Guo}}, \bibinfo {author} {\bibfnamefont {Y.}~\bibnamefont {Xu}}, \bibinfo {author} {\bibfnamefont {Y.}~\bibnamefont {Su}}, \bibinfo {author} {\bibfnamefont {E.}~\bibnamefont {Feng}}, \bibinfo {author} {\bibfnamefont {Y.}~\bibnamefont {Liu}}, \bibinfo {author} {\bibfnamefont {S.}~\bibnamefont {Zhou}}, \bibinfo {author} {\bibfnamefont {T.}~\bibnamefont {Ying}}, \bibinfo {author} {\bibfnamefont {S.}~\bibnamefont {Li}}, \bibinfo {author} {\bibfnamefont {Z.}~\bibnamefont {Wang}}, \bibinfo {author} {\bibfnamefont {G.}~\bibnamefont {Chen}},\ and\ \bibinfo {author} {\bibfnamefont {X.}~\bibnamefont {Chen}},\ }\href {https://doi.org/10.1103/PhysRevB.99.054421} {\bibfield  {journal} {\bibinfo  {journal} {Phys. Rev. B}\ }\textbf {\bibinfo {volume} {99}},\ \bibinfo {pages} {054421} (\bibinfo {year} {2019})}\BibitemShut {NoStop}%
\bibitem [{\citenamefont {Bordelon}\ \emph {et~al.}(2019)\citenamefont {Bordelon}, \citenamefont {Kenney}, \citenamefont {Liu}, \citenamefont {Hogan}, \citenamefont {Posthuma}, \citenamefont {Kavand}, \citenamefont {Lyu}, \citenamefont {Sherwin}, \citenamefont {Butch}, \citenamefont {Brown} \emph {et~al.}}]{Bordelon:NaYbO2}%
  \BibitemOpen
  \bibfield  {author} {\bibinfo {author} {\bibfnamefont {M.~M.}\ \bibnamefont {Bordelon}}, \bibinfo {author} {\bibfnamefont {E.}~\bibnamefont {Kenney}}, \bibinfo {author} {\bibfnamefont {C.}~\bibnamefont {Liu}}, \bibinfo {author} {\bibfnamefont {T.}~\bibnamefont {Hogan}}, \bibinfo {author} {\bibfnamefont {L.}~\bibnamefont {Posthuma}}, \bibinfo {author} {\bibfnamefont {M.}~\bibnamefont {Kavand}}, \bibinfo {author} {\bibfnamefont {Y.}~\bibnamefont {Lyu}}, \bibinfo {author} {\bibfnamefont {M.}~\bibnamefont {Sherwin}}, \bibinfo {author} {\bibfnamefont {N.~P.}\ \bibnamefont {Butch}}, \bibinfo {author} {\bibfnamefont {C.}~\bibnamefont {Brown}}, \emph {et~al.},\ }\href@noop {} {\bibfield  {journal} {\bibinfo  {journal} {Nature Physics}\ }\textbf {\bibinfo {volume} {15}},\ \bibinfo {pages} {1058} (\bibinfo {year} {2019})}\BibitemShut {NoStop}%
\bibitem [{\citenamefont {Ranjith}\ \emph {et~al.}(2019)\citenamefont {Ranjith}, \citenamefont {Luther}, \citenamefont {Reimann}, \citenamefont {Schmidt}, \citenamefont {Schlender}, \citenamefont {Sichelschmidt}, \citenamefont {Yasuoka}, \citenamefont {Strydom}, \citenamefont {Skourski}, \citenamefont {Wosnitza}, \citenamefont {K\"uhne}, \citenamefont {Doert},\ and\ \citenamefont {Baenitz}}]{Ranjith:NaYbSe2}%
  \BibitemOpen
  \bibfield  {author} {\bibinfo {author} {\bibfnamefont {K.~M.}\ \bibnamefont {Ranjith}}, \bibinfo {author} {\bibfnamefont {S.}~\bibnamefont {Luther}}, \bibinfo {author} {\bibfnamefont {T.}~\bibnamefont {Reimann}}, \bibinfo {author} {\bibfnamefont {B.}~\bibnamefont {Schmidt}}, \bibinfo {author} {\bibfnamefont {P.}~\bibnamefont {Schlender}}, \bibinfo {author} {\bibfnamefont {J.}~\bibnamefont {Sichelschmidt}}, \bibinfo {author} {\bibfnamefont {H.}~\bibnamefont {Yasuoka}}, \bibinfo {author} {\bibfnamefont {A.~M.}\ \bibnamefont {Strydom}}, \bibinfo {author} {\bibfnamefont {Y.}~\bibnamefont {Skourski}}, \bibinfo {author} {\bibfnamefont {J.}~\bibnamefont {Wosnitza}}, \bibinfo {author} {\bibfnamefont {H.}~\bibnamefont {K\"uhne}}, \bibinfo {author} {\bibfnamefont {T.}~\bibnamefont {Doert}},\ and\ \bibinfo {author} {\bibfnamefont {M.}~\bibnamefont {Baenitz}},\ }\href {https://doi.org/10.1103/PhysRevB.100.224417} {\bibfield  {journal} {\bibinfo  {journal} {Phys. Rev. B}\ }\textbf {\bibinfo {volume} {100}},\ \bibinfo
  {pages} {224417} (\bibinfo {year} {2019})}\BibitemShut {NoStop}%
\bibitem [{\citenamefont {Sugiura}\ and\ \citenamefont {Shimizu}(2013)}]{SugiuraShimizu:KHAF}%
  \BibitemOpen
  \bibfield  {author} {\bibinfo {author} {\bibfnamefont {S.}~\bibnamefont {Sugiura}}\ and\ \bibinfo {author} {\bibfnamefont {A.}~\bibnamefont {Shimizu}},\ }\href {https://doi.org/10.1103/PhysRevLett.111.010401} {\bibfield  {journal} {\bibinfo  {journal} {Phys. Rev. Lett.}\ }\textbf {\bibinfo {volume} {111}},\ \bibinfo {pages} {010401} (\bibinfo {year} {2013})}\BibitemShut {NoStop}%
\bibitem [{\citenamefont {Munehisa}(2014)}]{Munehisa:KHAF}%
  \BibitemOpen
  \bibfield  {author} {\bibinfo {author} {\bibfnamefont {T.}~\bibnamefont {Munehisa}},\ }\href {https://doi.org/10.4236/wjcmp.2014.43018} {\bibfield  {journal} {\bibinfo  {journal} {World Journal of Condensed Matter Physics}\ }\textbf {\bibinfo {volume} {04}},\ \bibinfo {pages} {134–140} (\bibinfo {year} {2014})}\BibitemShut {NoStop}%
\bibitem [{\citenamefont {Chen}\ \emph {et~al.}(2019)\citenamefont {Chen}, \citenamefont {Qu}, \citenamefont {Li}, \citenamefont {Chen}, \citenamefont {Gong}, \citenamefont {von Delft}, \citenamefont {Weichselbaum},\ and\ \citenamefont {Li}}]{ChenQu:THAF}%
  \BibitemOpen
  \bibfield  {author} {\bibinfo {author} {\bibfnamefont {L.}~\bibnamefont {Chen}}, \bibinfo {author} {\bibfnamefont {D.-W.}\ \bibnamefont {Qu}}, \bibinfo {author} {\bibfnamefont {H.}~\bibnamefont {Li}}, \bibinfo {author} {\bibfnamefont {B.-B.}\ \bibnamefont {Chen}}, \bibinfo {author} {\bibfnamefont {S.-S.}\ \bibnamefont {Gong}}, \bibinfo {author} {\bibfnamefont {J.}~\bibnamefont {von Delft}}, \bibinfo {author} {\bibfnamefont {A.}~\bibnamefont {Weichselbaum}},\ and\ \bibinfo {author} {\bibfnamefont {W.}~\bibnamefont {Li}},\ }\href {https://doi.org/10.1103/PhysRevB.99.140404} {\bibfield  {journal} {\bibinfo  {journal} {Phys. Rev. B}\ }\textbf {\bibinfo {volume} {99}},\ \bibinfo {pages} {140404} (\bibinfo {year} {2019})}\BibitemShut {NoStop}%
\bibitem [{\citenamefont {Prelov\ifmmode~\check{s}\else \v{s}\fi{}ek}\ and\ \citenamefont {Kokalj}(2018)}]{PrelovsekKokalj:THAF}%
  \BibitemOpen
  \bibfield  {author} {\bibinfo {author} {\bibfnamefont {P.}~\bibnamefont {Prelov\ifmmode~\check{s}\else \v{s}\fi{}ek}}\ and\ \bibinfo {author} {\bibfnamefont {J.}~\bibnamefont {Kokalj}},\ }\href {https://doi.org/10.1103/PhysRevB.98.035107} {\bibfield  {journal} {\bibinfo  {journal} {Phys. Rev. B}\ }\textbf {\bibinfo {volume} {98}},\ \bibinfo {pages} {035107} (\bibinfo {year} {2018})}\BibitemShut {NoStop}%
\bibitem [{\citenamefont {Schnack}\ \emph {et~al.}(2018)\citenamefont {Schnack}, \citenamefont {Schulenburg},\ and\ \citenamefont {Richter}}]{SchnackSchulenberg:KHAF}%
  \BibitemOpen
  \bibfield  {author} {\bibinfo {author} {\bibfnamefont {J.}~\bibnamefont {Schnack}}, \bibinfo {author} {\bibfnamefont {J.}~\bibnamefont {Schulenburg}},\ and\ \bibinfo {author} {\bibfnamefont {J.}~\bibnamefont {Richter}},\ }\href {https://doi.org/10.1103/PhysRevB.98.094423} {\bibfield  {journal} {\bibinfo  {journal} {Phys. Rev. B}\ }\textbf {\bibinfo {volume} {98}},\ \bibinfo {pages} {094423} (\bibinfo {year} {2018})}\BibitemShut {NoStop}%
\bibitem [{\citenamefont {Gonzalez}\ \emph {et~al.}(2022)\citenamefont {Gonzalez}, \citenamefont {Bernu}, \citenamefont {Pierre},\ and\ \citenamefont {Messio}}]{Gonzalez:THAF}%
  \BibitemOpen
  \bibfield  {author} {\bibinfo {author} {\bibfnamefont {M.~G.}\ \bibnamefont {Gonzalez}}, \bibinfo {author} {\bibfnamefont {B.}~\bibnamefont {Bernu}}, \bibinfo {author} {\bibfnamefont {L.}~\bibnamefont {Pierre}},\ and\ \bibinfo {author} {\bibfnamefont {L.}~\bibnamefont {Messio}},\ }\href {https://doi.org/10.21468/SciPostPhys.12.3.112} {\bibfield  {journal} {\bibinfo  {journal} {SciPost Phys.}\ }\textbf {\bibinfo {volume} {12}},\ \bibinfo {pages} {112} (\bibinfo {year} {2022})}\BibitemShut {NoStop}%
\bibitem [{\citenamefont {Landau}(1983)}]{Landau:Vortices}%
  \BibitemOpen
  \bibfield  {author} {\bibinfo {author} {\bibfnamefont {D.~P.}\ \bibnamefont {Landau}},\ }\href {https://doi.org/10.1103/PhysRevB.27.5604} {\bibfield  {journal} {\bibinfo  {journal} {Phys. Rev. B}\ }\textbf {\bibinfo {volume} {27}},\ \bibinfo {pages} {5604} (\bibinfo {year} {1983})}\BibitemShut {NoStop}%
\bibitem [{\citenamefont {Moore}\ \emph {et~al.}(1999)\citenamefont {Moore}, \citenamefont {Nordahl}, \citenamefont {Minar},\ and\ \citenamefont {Shalizi}}]{Moore:LogInteractions}%
  \BibitemOpen
  \bibfield  {author} {\bibinfo {author} {\bibfnamefont {C.}~\bibnamefont {Moore}}, \bibinfo {author} {\bibfnamefont {M.~G.}\ \bibnamefont {Nordahl}}, \bibinfo {author} {\bibfnamefont {N.}~\bibnamefont {Minar}},\ and\ \bibinfo {author} {\bibfnamefont {C.~R.}\ \bibnamefont {Shalizi}},\ }\href {https://doi.org/10.1103/PhysRevE.60.5344} {\bibfield  {journal} {\bibinfo  {journal} {Phys. Rev. E}\ }\textbf {\bibinfo {volume} {60}},\ \bibinfo {pages} {5344} (\bibinfo {year} {1999})}\BibitemShut {NoStop}%
\end{thebibliography}%


\begin{thebibliography}{20}%
\makeatletter
\providecommand \@ifxundefined [1]{%
 \@ifx{#1\undefined}
}%
\providecommand \@ifnum [1]{%
 \ifnum #1\expandafter \@firstoftwo
 \else \expandafter \@secondoftwo
 \fi
}%
\providecommand \@ifx [1]{%
 \ifx #1\expandafter \@firstoftwo
 \else \expandafter \@secondoftwo
 \fi
}%
\providecommand \natexlab [1]{#1}%
\providecommand \enquote  [1]{``#1''}%
\providecommand \bibnamefont  [1]{#1}%
\providecommand \bibfnamefont [1]{#1}%
\providecommand \citenamefont [1]{#1}%
\providecommand \href@noop [0]{\@secondoftwo}%
\providecommand \href [0]{\begingroup \@sanitize@url \@href}%
\providecommand \@href[1]{\@@startlink{#1}\@@href}%
\providecommand \@@href[1]{\endgroup#1\@@endlink}%
\providecommand \@sanitize@url [0]{\catcode `\\12\catcode `\$12\catcode `\&12\catcode `\#12\catcode `\^12\catcode `\_12\catcode `\%12\relax}%
\providecommand \@@startlink[1]{}%
\providecommand \@@endlink[0]{}%
\providecommand \url  [0]{\begingroup\@sanitize@url \@url }%
\providecommand \@url [1]{\endgroup\@href {#1}{\urlprefix }}%
\providecommand \urlprefix  [0]{URL }%
\providecommand \Eprint [0]{\href }%
\providecommand \doibase [0]{https://doi.org/}%
\providecommand \selectlanguage [0]{\@gobble}%
\providecommand \bibinfo  [0]{\@secondoftwo}%
\providecommand \bibfield  [0]{\@secondoftwo}%
\providecommand \translation [1]{[#1]}%
\providecommand \BibitemOpen [0]{}%
\providecommand \bibitemStop [0]{}%
\providecommand \bibitemNoStop [0]{.\EOS\space}%
\providecommand \EOS [0]{\spacefactor3000\relax}%
\providecommand \BibitemShut  [1]{\csname bibitem#1\endcsname}%
\let\auto@bib@innerbib\@empty
\bibitem [{\citenamefont {Kac}\ and\ \citenamefont {Ward}(1952)}]{kac1952combinatorial}%
  \BibitemOpen
  \bibfield  {author} {\bibinfo {author} {\bibfnamefont {M.}~\bibnamefont {Kac}}\ and\ \bibinfo {author} {\bibfnamefont {J.~C.}\ \bibnamefont {Ward}},\ }\href {https://doi.org/10.1103/PhysRev.88.1332} {\bibfield  {journal} {\bibinfo  {journal} {Phys. Rev.}\ }\textbf {\bibinfo {volume} {88}},\ \bibinfo {pages} {1332} (\bibinfo {year} {1952})}\BibitemShut {NoStop}%
\bibitem [{\citenamefont {Landau}\ and\ \citenamefont {Lifshitz}(2013)}]{Landafshitz5}%
  \BibitemOpen
  \bibfield  {author} {\bibinfo {author} {\bibfnamefont {L.~D.}\ \bibnamefont {Landau}}\ and\ \bibinfo {author} {\bibfnamefont {E.~M.}\ \bibnamefont {Lifshitz}},\ }\href@noop {} {\emph {\bibinfo {title} {Statistical Physics: Volume 5}}},\ Vol.~\bibinfo {volume} {5}\ (\bibinfo  {publisher} {Elsevier},\ \bibinfo {year} {2013})\BibitemShut {NoStop}%
\bibitem [{\citenamefont {Wannier}(1950)}]{Wannier:Ising}%
  \BibitemOpen
  \bibfield  {author} {\bibinfo {author} {\bibfnamefont {G.~H.}\ \bibnamefont {Wannier}},\ }\href {https://doi.org/10.1103/PhysRev.79.357} {\bibfield  {journal} {\bibinfo  {journal} {Phys. Rev.}\ }\textbf {\bibinfo {volume} {79}},\ \bibinfo {pages} {357} (\bibinfo {year} {1950})}\BibitemShut {NoStop}%
\bibitem [{\citenamefont {Wannier}(1973)}]{Wannier:erratum}%
  \BibitemOpen
  \bibfield  {author} {\bibinfo {author} {\bibfnamefont {G.~H.}\ \bibnamefont {Wannier}},\ }\href {https://doi.org/10.1103/PhysRevB.7.5017} {\bibfield  {journal} {\bibinfo  {journal} {Phys. Rev. B}\ }\textbf {\bibinfo {volume} {7}},\ \bibinfo {pages} {5017} (\bibinfo {year} {1973})}\BibitemShut {NoStop}%
\bibitem [{\citenamefont {Landau}(1983)}]{Landau:Vortices}%
  \BibitemOpen
  \bibfield  {author} {\bibinfo {author} {\bibfnamefont {D.~P.}\ \bibnamefont {Landau}},\ }\href {https://doi.org/10.1103/PhysRevB.27.5604} {\bibfield  {journal} {\bibinfo  {journal} {Phys. Rev. B}\ }\textbf {\bibinfo {volume} {27}},\ \bibinfo {pages} {5604} (\bibinfo {year} {1983})}\BibitemShut {NoStop}%
\bibitem [{\citenamefont {Moore}\ \emph {et~al.}(1999)\citenamefont {Moore}, \citenamefont {Nordahl}, \citenamefont {Minar},\ and\ \citenamefont {Shalizi}}]{Moore:LogInteractions}%
  \BibitemOpen
  \bibfield  {author} {\bibinfo {author} {\bibfnamefont {C.}~\bibnamefont {Moore}}, \bibinfo {author} {\bibfnamefont {M.~G.}\ \bibnamefont {Nordahl}}, \bibinfo {author} {\bibfnamefont {N.}~\bibnamefont {Minar}},\ and\ \bibinfo {author} {\bibfnamefont {C.~R.}\ \bibnamefont {Shalizi}},\ }\href {https://doi.org/10.1103/PhysRevE.60.5344} {\bibfield  {journal} {\bibinfo  {journal} {Phys. Rev. E}\ }\textbf {\bibinfo {volume} {60}},\ \bibinfo {pages} {5344} (\bibinfo {year} {1999})}\BibitemShut {NoStop}%
\bibitem [{\citenamefont {Chalker}(2017)}]{Chalker:notes}%
  \BibitemOpen
  \bibfield  {author} {\bibinfo {author} {\bibfnamefont {J.~T.}\ \bibnamefont {Chalker}},\ }\href@noop {} {\bibfield  {journal} {\bibinfo  {journal} {Topological Aspects of Condensed Matter Physics}\ ,\ \bibinfo {pages} {123}} (\bibinfo {year} {2017})}\BibitemShut {NoStop}%
\bibitem [{\citenamefont {Popp}\ \emph {et~al.}(2025)\citenamefont {Popp}, \citenamefont {Ramirez},\ and\ \citenamefont {Syzranov}}]{PoppRamirezSyzranov}%
  \BibitemOpen
  \bibfield  {author} {\bibinfo {author} {\bibfnamefont {P.}~\bibnamefont {Popp}}, \bibinfo {author} {\bibfnamefont {A.~P.}\ \bibnamefont {Ramirez}},\ and\ \bibinfo {author} {\bibfnamefont {S.}~\bibnamefont {Syzranov}},\ }\href {https://doi.org/10.1103/PhysRevLett.134.226701} {\bibfield  {journal} {\bibinfo  {journal} {Phys. Rev. Lett.}\ }\textbf {\bibinfo {volume} {134}},\ \bibinfo {pages} {226701} (\bibinfo {year} {2025})}\BibitemShut {NoStop}%
\bibitem [{\citenamefont {Ramirez}\ and\ \citenamefont {Syzranov}(2025)}]{RamirezSyzranov:review}%
  \BibitemOpen
  \bibfield  {author} {\bibinfo {author} {\bibfnamefont {A.~P.}\ \bibnamefont {Ramirez}}\ and\ \bibinfo {author} {\bibfnamefont {S.~V.}\ \bibnamefont {Syzranov}},\ }\href {https://doi.org/10.1039/D4MA00914B} {\bibfield  {journal} {\bibinfo  {journal} {Mater. Adv.}\ }\textbf {\bibinfo {volume} {6}},\ \bibinfo {pages} {1213} (\bibinfo {year} {2025})}\BibitemShut {NoStop}%
\bibitem [{\citenamefont {Prelov\ifmmode~\check{s}\else \v{s}\fi{}ek}\ and\ \citenamefont {Kokalj}(2018)}]{PrelovsekKokalj:THAF}%
  \BibitemOpen
  \bibfield  {author} {\bibinfo {author} {\bibfnamefont {P.}~\bibnamefont {Prelov\ifmmode~\check{s}\else \v{s}\fi{}ek}}\ and\ \bibinfo {author} {\bibfnamefont {J.}~\bibnamefont {Kokalj}},\ }\href {https://doi.org/10.1103/PhysRevB.98.035107} {\bibfield  {journal} {\bibinfo  {journal} {Phys. Rev. B}\ }\textbf {\bibinfo {volume} {98}},\ \bibinfo {pages} {035107} (\bibinfo {year} {2018})}\BibitemShut {NoStop}%
\bibitem [{\citenamefont {Chen}\ \emph {et~al.}(2019)\citenamefont {Chen}, \citenamefont {Qu}, \citenamefont {Li}, \citenamefont {Chen}, \citenamefont {Gong}, \citenamefont {von Delft}, \citenamefont {Weichselbaum},\ and\ \citenamefont {Li}}]{ChenQu:THAF}%
  \BibitemOpen
  \bibfield  {author} {\bibinfo {author} {\bibfnamefont {L.}~\bibnamefont {Chen}}, \bibinfo {author} {\bibfnamefont {D.-W.}\ \bibnamefont {Qu}}, \bibinfo {author} {\bibfnamefont {H.}~\bibnamefont {Li}}, \bibinfo {author} {\bibfnamefont {B.-B.}\ \bibnamefont {Chen}}, \bibinfo {author} {\bibfnamefont {S.-S.}\ \bibnamefont {Gong}}, \bibinfo {author} {\bibfnamefont {J.}~\bibnamefont {von Delft}}, \bibinfo {author} {\bibfnamefont {A.}~\bibnamefont {Weichselbaum}},\ and\ \bibinfo {author} {\bibfnamefont {W.}~\bibnamefont {Li}},\ }\href {https://doi.org/10.1103/PhysRevB.99.140404} {\bibfield  {journal} {\bibinfo  {journal} {Phys. Rev. B}\ }\textbf {\bibinfo {volume} {99}},\ \bibinfo {pages} {140404} (\bibinfo {year} {2019})}\BibitemShut {NoStop}%
\bibitem [{\citenamefont {Seki}\ and\ \citenamefont {Yunoki}(2020)}]{SekiYunoki:RingExchange}%
  \BibitemOpen
  \bibfield  {author} {\bibinfo {author} {\bibfnamefont {K.}~\bibnamefont {Seki}}\ and\ \bibinfo {author} {\bibfnamefont {S.}~\bibnamefont {Yunoki}},\ }\href {https://doi.org/10.1103/PhysRevB.101.235115} {\bibfield  {journal} {\bibinfo  {journal} {Phys. Rev. B}\ }\textbf {\bibinfo {volume} {101}},\ \bibinfo {pages} {235115} (\bibinfo {year} {2020})}\BibitemShut {NoStop}%
\bibitem [{\citenamefont {{Ulaga}}\ \emph {et~al.}(2024)\citenamefont {{Ulaga}}, \citenamefont {{Kokalj}},\ and\ \citenamefont {{Prelov{\v{s}}ek}}}]{UlagaPrelovsek:latest}%
  \BibitemOpen
  \bibfield  {author} {\bibinfo {author} {\bibfnamefont {M.}~\bibnamefont {{Ulaga}}}, \bibinfo {author} {\bibfnamefont {J.}~\bibnamefont {{Kokalj}}},\ and\ \bibinfo {author} {\bibfnamefont {P.}~\bibnamefont {{Prelov{\v{s}}ek}}},\ }\href {https://doi.org/10.48550/arXiv.2408.05034} {\bibfield  {journal} {\bibinfo  {journal} {arXiv e-prints}\ ,\ \bibinfo {eid} {arXiv:2408.05034}} (\bibinfo {year} {2024})},\ \Eprint {https://arxiv.org/abs/2408.05034} {arXiv:2408.05034 [cond-mat.str-el]} \BibitemShut {NoStop}%
\bibitem [{\citenamefont {Elser}(1989)}]{Elser:KHAF}%
  \BibitemOpen
  \bibfield  {author} {\bibinfo {author} {\bibfnamefont {V.}~\bibnamefont {Elser}},\ }\href {https://doi.org/10.1103/PhysRevLett.62.2405} {\bibfield  {journal} {\bibinfo  {journal} {Phys. Rev. Lett.}\ }\textbf {\bibinfo {volume} {62}},\ \bibinfo {pages} {2405} (\bibinfo {year} {1989})}\BibitemShut {NoStop}%
\bibitem [{\citenamefont {Sugiura}\ and\ \citenamefont {Shimizu}(2013)}]{SugiuraShimizu:KHAF}%
  \BibitemOpen
  \bibfield  {author} {\bibinfo {author} {\bibfnamefont {S.}~\bibnamefont {Sugiura}}\ and\ \bibinfo {author} {\bibfnamefont {A.}~\bibnamefont {Shimizu}},\ }\href {https://doi.org/10.1103/PhysRevLett.111.010401} {\bibfield  {journal} {\bibinfo  {journal} {Phys. Rev. Lett.}\ }\textbf {\bibinfo {volume} {111}},\ \bibinfo {pages} {010401} (\bibinfo {year} {2013})}\BibitemShut {NoStop}%
\bibitem [{\citenamefont {Shimokawa}\ and\ \citenamefont {Kawamura}(2016)}]{shimokawa2016finite}%
  \BibitemOpen
  \bibfield  {author} {\bibinfo {author} {\bibfnamefont {T.}~\bibnamefont {Shimokawa}}\ and\ \bibinfo {author} {\bibfnamefont {H.}~\bibnamefont {Kawamura}},\ }\href {https://doi.org/10.7566/JPSJ.85.113702} {\bibfield  {journal} {\bibinfo  {journal} {Journal of the Physical Society of Japan}\ }\textbf {\bibinfo {volume} {85}},\ \bibinfo {pages} {113702} (\bibinfo {year} {2016})}\BibitemShut {NoStop}%
\bibitem [{\citenamefont {Schnack}\ \emph {et~al.}(2018)\citenamefont {Schnack}, \citenamefont {Schulenburg},\ and\ \citenamefont {Richter}}]{SchnackSchulenberg:KHAF}%
  \BibitemOpen
  \bibfield  {author} {\bibinfo {author} {\bibfnamefont {J.}~\bibnamefont {Schnack}}, \bibinfo {author} {\bibfnamefont {J.}~\bibnamefont {Schulenburg}},\ and\ \bibinfo {author} {\bibfnamefont {J.}~\bibnamefont {Richter}},\ }\href {https://doi.org/10.1103/PhysRevB.98.094423} {\bibfield  {journal} {\bibinfo  {journal} {Phys. Rev. B}\ }\textbf {\bibinfo {volume} {98}},\ \bibinfo {pages} {094423} (\bibinfo {year} {2018})}\BibitemShut {NoStop}%
\bibitem [{\citenamefont {Li}\ \emph {et~al.}(2019)\citenamefont {Li}, \citenamefont {Jin}, \citenamefont {Guo}, \citenamefont {Xu}, \citenamefont {Su}, \citenamefont {Feng}, \citenamefont {Liu}, \citenamefont {Zhou}, \citenamefont {Ying}, \citenamefont {Li}, \citenamefont {Wang}, \citenamefont {Chen},\ and\ \citenamefont {Chen}}]{Li:S2TAF}%
  \BibitemOpen
  \bibfield  {author} {\bibinfo {author} {\bibfnamefont {K.}~\bibnamefont {Li}}, \bibinfo {author} {\bibfnamefont {S.}~\bibnamefont {Jin}}, \bibinfo {author} {\bibfnamefont {J.}~\bibnamefont {Guo}}, \bibinfo {author} {\bibfnamefont {Y.}~\bibnamefont {Xu}}, \bibinfo {author} {\bibfnamefont {Y.}~\bibnamefont {Su}}, \bibinfo {author} {\bibfnamefont {E.}~\bibnamefont {Feng}}, \bibinfo {author} {\bibfnamefont {Y.}~\bibnamefont {Liu}}, \bibinfo {author} {\bibfnamefont {S.}~\bibnamefont {Zhou}}, \bibinfo {author} {\bibfnamefont {T.}~\bibnamefont {Ying}}, \bibinfo {author} {\bibfnamefont {S.}~\bibnamefont {Li}}, \bibinfo {author} {\bibfnamefont {Z.}~\bibnamefont {Wang}}, \bibinfo {author} {\bibfnamefont {G.}~\bibnamefont {Chen}},\ and\ \bibinfo {author} {\bibfnamefont {X.}~\bibnamefont {Chen}},\ }\href {https://doi.org/10.1103/PhysRevB.99.054421} {\bibfield  {journal} {\bibinfo  {journal} {Phys. Rev. B}\ }\textbf {\bibinfo {volume} {99}},\ \bibinfo {pages} {054421} (\bibinfo {year} {2019})}\BibitemShut {NoStop}%
\bibitem [{\citenamefont {Bordelon}\ \emph {et~al.}(2019)\citenamefont {Bordelon}, \citenamefont {Kenney}, \citenamefont {Liu}, \citenamefont {Hogan}, \citenamefont {Posthuma}, \citenamefont {Kavand}, \citenamefont {Lyu}, \citenamefont {Sherwin}, \citenamefont {Butch}, \citenamefont {Brown} \emph {et~al.}}]{Bordelon:NaYbO2}%
  \BibitemOpen
  \bibfield  {author} {\bibinfo {author} {\bibfnamefont {M.~M.}\ \bibnamefont {Bordelon}}, \bibinfo {author} {\bibfnamefont {E.}~\bibnamefont {Kenney}}, \bibinfo {author} {\bibfnamefont {C.}~\bibnamefont {Liu}}, \bibinfo {author} {\bibfnamefont {T.}~\bibnamefont {Hogan}}, \bibinfo {author} {\bibfnamefont {L.}~\bibnamefont {Posthuma}}, \bibinfo {author} {\bibfnamefont {M.}~\bibnamefont {Kavand}}, \bibinfo {author} {\bibfnamefont {Y.}~\bibnamefont {Lyu}}, \bibinfo {author} {\bibfnamefont {M.}~\bibnamefont {Sherwin}}, \bibinfo {author} {\bibfnamefont {N.~P.}\ \bibnamefont {Butch}}, \bibinfo {author} {\bibfnamefont {C.}~\bibnamefont {Brown}}, \emph {et~al.},\ }\href@noop {} {\bibfield  {journal} {\bibinfo  {journal} {Nature Physics}\ }\textbf {\bibinfo {volume} {15}},\ \bibinfo {pages} {1058} (\bibinfo {year} {2019})}\BibitemShut {NoStop}%
\bibitem [{\citenamefont {Ranjith}\ \emph {et~al.}(2019)\citenamefont {Ranjith}, \citenamefont {Luther}, \citenamefont {Reimann}, \citenamefont {Schmidt}, \citenamefont {Schlender}, \citenamefont {Sichelschmidt}, \citenamefont {Yasuoka}, \citenamefont {Strydom}, \citenamefont {Skourski}, \citenamefont {Wosnitza}, \citenamefont {K\"uhne}, \citenamefont {Doert},\ and\ \citenamefont {Baenitz}}]{Ranjith:NaYbSe2}%
  \BibitemOpen
  \bibfield  {author} {\bibinfo {author} {\bibfnamefont {K.~M.}\ \bibnamefont {Ranjith}}, \bibinfo {author} {\bibfnamefont {S.}~\bibnamefont {Luther}}, \bibinfo {author} {\bibfnamefont {T.}~\bibnamefont {Reimann}}, \bibinfo {author} {\bibfnamefont {B.}~\bibnamefont {Schmidt}}, \bibinfo {author} {\bibfnamefont {P.}~\bibnamefont {Schlender}}, \bibinfo {author} {\bibfnamefont {J.}~\bibnamefont {Sichelschmidt}}, \bibinfo {author} {\bibfnamefont {H.}~\bibnamefont {Yasuoka}}, \bibinfo {author} {\bibfnamefont {A.~M.}\ \bibnamefont {Strydom}}, \bibinfo {author} {\bibfnamefont {Y.}~\bibnamefont {Skourski}}, \bibinfo {author} {\bibfnamefont {J.}~\bibnamefont {Wosnitza}}, \bibinfo {author} {\bibfnamefont {H.}~\bibnamefont {K\"uhne}}, \bibinfo {author} {\bibfnamefont {T.}~\bibnamefont {Doert}},\ and\ \bibinfo {author} {\bibfnamefont {M.}~\bibnamefont {Baenitz}},\ }\href {https://doi.org/10.1103/PhysRevB.100.224417} {\bibfield  {journal} {\bibinfo  {journal} {Phys. Rev. B}\ }\textbf {\bibinfo {volume} {100}},\ \bibinfo
  {pages} {224417} (\bibinfo {year} {2019})}\BibitemShut {NoStop}%
\end{thebibliography}%


\begin{thebibliography}{0}%
\makeatletter
\providecommand \@ifxundefined [1]{%
 \@ifx{#1\undefined}
}%
\providecommand \@ifnum [1]{%
 \ifnum #1\expandafter \@firstoftwo
 \else \expandafter \@secondoftwo
 \fi
}%
\providecommand \@ifx [1]{%
 \ifx #1\expandafter \@firstoftwo
 \else \expandafter \@secondoftwo
 \fi
}%
\providecommand \natexlab [1]{#1}%
\providecommand \enquote  [1]{``#1''}%
\providecommand \bibnamefont  [1]{#1}%
\providecommand \bibfnamefont [1]{#1}%
\providecommand \citenamefont [1]{#1}%
\providecommand \href@noop [0]{\@secondoftwo}%
\providecommand \href [0]{\begingroup \@sanitize@url \@href}%
\providecommand \@href[1]{\@@startlink{#1}\@@href}%
\providecommand \@@href[1]{\endgroup#1\@@endlink}%
\providecommand \@sanitize@url [0]{\catcode `\\12\catcode `\$12\catcode `\&12\catcode `\#12\catcode `\^12\catcode `\_12\catcode `\%12\relax}%
\providecommand \@@startlink[1]{}%
\providecommand \@@endlink[0]{}%
\providecommand \url  [0]{\begingroup\@sanitize@url \@url }%
\providecommand \@url [1]{\endgroup\@href {#1}{\urlprefix }}%
\providecommand \urlprefix  [0]{URL }%
\providecommand \Eprint [0]{\href }%
\providecommand \doibase [0]{https://doi.org/}%
\providecommand \selectlanguage [0]{\@gobble}%
\providecommand \bibinfo  [0]{\@secondoftwo}%
\providecommand \bibfield  [0]{\@secondoftwo}%
\providecommand \translation [1]{[#1]}%
\providecommand \BibitemOpen [0]{}%
\providecommand \bibitemStop [0]{}%
\providecommand \bibitemNoStop [0]{.\EOS\space}%
\providecommand \EOS [0]{\spacefactor3000\relax}%
\providecommand \BibitemShut  [1]{\csname bibitem#1\endcsname}%
\let\auto@bib@innerbib\@empty
\end{thebibliography}%
\end{document}